\documentclass[ prd, twocolumn, nofootinbib, floatfix]{revtex4-1} 

\usepackage{mathptmx}
\DeclareMathAlphabet{\mathcal}{OMS}{cmsy}{m}{n}

\usepackage{dsfont}
\usepackage[mathscr]{euscript}
\usepackage{amsmath}
\usepackage{graphicx}
\usepackage{dcolumn}
\usepackage{bm}
\usepackage{epsfig}
\usepackage{amssymb,latexsym,mathrsfs}
\usepackage{graphicx}
\usepackage{color}
\usepackage{hyperref}
\usepackage{float}
\usepackage{diagbox}
  \usepackage{cancel}

\definecolor{vividviolet}{rgb}{0.62, 0.0, 1.0}
\definecolor{amaranth}{rgb}{0.9, 0.17, 0.31}
\definecolor{palatinateblue}{rgb}{0.15, 0.23, 0.89}
\definecolor{brightpink}{rgb}{1.0, 0.0, 0.5}
\definecolor{cornflowerblue}{rgb}{0.39, 0.58, 0.93}
\definecolor{deepcarminepink}{rgb}{0.94, 0.19, 0.22}
\definecolor{radicalred}{rgb}{1.0, 0.21, 0.37}
\definecolor{blueblue}{RGB}{21,47,181}
\definecolor{greengreen}{RGB}{65,166,16}

\usepackage{xtab,afterpage}
\usepackage{hyperref}
\usepackage{setspace,calc}

\hypersetup{ linktoc=all,
    colorlinks, linkcolor={blueblue},
    citecolor={greengreen}, urlcolor={blueblue}
}

\newcommand{\be}{\begin{equation}}
\newcommand{\ee}{\end{equation}}
\newcommand{\bs}{\begin{split}} 
\newcommand{\bea}{\begin{eqnarray}}
\newcommand{\eea}{\end{eqnarray}}
\newcommand{\infint}{\int_{-\infty }^{\infty }}

\newcommand{\D}{\mathrm{d}}
\newsavebox{\myhbar}

\begin{document}

\title{Entanglement amplification between superposed detectors in flat and curved spacetimes}

\author{Joshua Foo${}^{1}$}
\email{joshua.foo@uqconnect.edu.au}
\author{Robert B. Mann${}^{2,3}$}
\email{rbmann@uwaterloo.ca}
\author{Magdalena Zych${}^{4}$}
\email{m.zych@uq.edu.au}
\affiliation{${}^1$Centre for Quantum Computation \& Communication Technology, School of Mathematics \& Physics,
The University of Queensland, St.~Lucia, Queensland, 4072, Australia\\
${}^2$Department of Physics and Astronomy, University of Waterloo, Waterloo, Ontario, Canada, N2L 3G1\\
${}^3$Perimeter Institute, 31 Caroline St., Waterloo, Ontario, N2L 2Y5, Canada
\\
${}^4$Centre for Engineered Quantum Systems, School of Mathematics and Physics, The University of Queensland, St. Lucia, Queensland, 4072, Australia}

\begin{abstract} 
We consider an entanglement harvesting protocol between two Unruh-deWitt detectors in quantum superpositions of static trajectories in the static de Sitter and thermal Minkowski spacetimes. We demonstrate for the first time that the spatial superposition of each detector's path allows entanglement to be harvested from the quantum field in regimes where it would be otherwise impossible for detectors on classical trajectories. Surprisingly, for detectors on sufficiently delocalised trajectories in a thermal bath, the amount of harvested entanglement grows with the temperature of the field, violating a no-go theorem derived by Simidzija et.\ al.\ (Phys.\ Rev.\ D \textbf{98}, 085007). We also discover interesting effects for the mutual information harvesting, which depends sensitively on the nonlocal correlations between the superposed trajectories between the paths of the respective detectors. 
\end{abstract}

\date{\today}

\maketitle 

\section{Introduction}
Entanglement describes the non-classical correlations that can exist within quantum systems, and is a fundamental property of the vacuum state in relativistic quantum field theory (QFT). In this regard, it underpins several important phenomena including the Unruh effect \cite{unruh1976notes} and Hawking radiation \cite{hawking1974black}, while from an information-theoretic perspective, it can be utilised as a physical resource which enables non-classical protocols such as quantum teleportation \cite{Koga:2018the,footeleporter}. 

Extracting entanglement from the vacuum state of quantum fields is a well-studied problem in the literature. Following the pioneering work of Valentini \cite{Valentini:1989fx} and Reznik et.\ al.\ \cite{Reznik:2002fz,reznik2005violating} emerged the field of \textit{entanglement harvesting}, which studies how initially uncorrelated quantum systems, interacting locally with the ambient background field, can become entangled. Entanglement harvesting has been shown to unveil novel information about the structure of entanglement in quantum fields and its dependence upon spacetime dimensionality \cite{pozas2015harvesting}, curvature \cite{ver2009entangling,henderson2018harvesting,kukita2017entanglement,ngAdS}, topology \cite{Martin-Martinez:2015qwa}, and the presence of horizons \cite{Cong:2020nec}. It has also been studied in experimentally feasible settings, see for example \cite{pozashydrogen}. 

In the studies mentioned, it is common to model the detectors using idealised, two-level systems (such as an atom) whose coupling to the ambient scalar field is a simple approximation to the light-matter interaction. This model, known as the Unruh-deWitt detector \cite{unruh1984happens}, has been recently applied to the problem of entanglement harvesting in the presence of indefinite causal order \cite{henderson2020quantum}. More specifically, the authors of \cite{henderson2020quantum} consider detectors whose interaction with the field occurs in a quantum-controlled superposition of temporal orders. Intriguingly, it was shown that the presence of temporal superposition enhances the ability of stationary detectors to extract entanglement from the Minkowski vacuum state. In particular, the no-go theorem of \cite{simidzija2018general}, which states that UdW detectors with Dirac-delta switching functions and arbitrary spatial profiles and detector-field couplings cannot harvest entanglement, was shown to be violated using the temporal superposition.

In this paper, we apply the quantum-controlled UdW detector model to entanglement harvesting in two well-known spacetimes; the static patch of de Sitter spacetime, and thermal Minkowski spacetime (i.e.\ flat Minkowski spacetime with a finite-temperature quantum field). We show that the presence of superposition -- not only the temporal superposition of the detector's interaction with the field, but also for detectors travelling in \textit{spatial superpositions} of spacetime trajectories -- enhances the ability of such detectors to harvest entanglement, compared to analogous scenarios with detectors on single trajectories. The presence of the superposition introduces quantum interference between the different field regions that each detector interacts with, suppressing the local excitations that they individually perceive. This typically allows the detectors to extract entanglement in regimes where it would be otherwise impossible for detectors travelling on classical trajectories. In fact, we demonstrate a violation of the no-go theorem derived by Simidzija et.\ al.\ in \cite{simidzijathermalstates}, which asserts that the entanglement harvested by detectors interacting with a thermal field state will \textit{decrease monotonically} with the temperature of the field. Since the theorem was derived under the tacit assumption of classical detector trajectories, our result shows that standard intuition (namely that higher temperatures detrimentally affect quantum entanglement) fails when extending the motion of detectors to include superpositions.
We also study the harvesting of mutual information, which measures the total amount of correlations between the detectors including classical correlations and non-entanglement quantum correlations. Interestingly, we discover that mutual information harvesting is \textit{inhibited} by the presence of the aforementioned quantum interference effects. 

Our paper is outlined as follows: in Sec.\ \ref{sec:qcudw}, we review the Unruh-deWitt detector model, including its extension to quantum-controlled superpositions of trajectories. We calculate the reduced density matrix of the bipartite detector system in Sec.\ \ref{sec:eh} and introduce the measures of entanglement we utilise in our analysis. In Sec.\ \ref{sec:desitter}, we analyse the entanglement harvesting protocol for detectors in superpositions of trajectories in the static patch of de Sitter spacetime. In Sec.\ \ref{thermalMinkowski}, we study quantum-controlled entanglement harvesting with finite-temperature fields in Minkowski spacetime. We summarise our results in Sec.\ \ref{sec:conc} and outline possibilities for future research. Throughout this paper, we adopt natural units, $G = c = \hslash = k_B = 1$. 

\section{Unruh-deWitt detectors in superpositions of trajectories}\label{sec:qcudw}

The standard UdW detector model considers a two-level system, whose internal states $|g\rangle, |e\rangle$ with energy gap $\Omega$, couple to the real, massless scalar field $\hat{\Phi}(\mathsf{x}(\tau))$ initially in the state $|\psi \rangle$, along the worldline $\mathsf{x}(\tau)$. The detector-field coupling is described via the interaction Hamiltonian, 
\begin{align}
    \hat{H}_\text{int.} &= \lambda \eta(\tau ) \sigma(\tau ) \hat{\Phi}(\mathsf{x}(\tau ) ) 
\end{align}
where $\lambda$ is a weak coupling constant, $\eta(\tau)$ is a switching function, $\sigma(\tau) = |e\rangle\langle g|e^{i\Omega\tau} + \text{H.c}$ is the interaction picture Pauli operator, acting on the detector Hilbert space. To model the detector as travelling in a superposition of trajectories, we follow recent works \cite{foo2020unruhdewitt,Barbado:2020snx,foo2020thermality,Foo:2020jmi} by introducing a control degree of freedom, $|\chi\rangle$, to the initial state of the detector, so that the initial detector-field state is described by 
\begin{align}
    |\Psi\rangle_{CFD} &= |\chi\rangle |\psi \rangle|g\rangle \qquad \text{where} \qquad |\chi\rangle  =\frac{1}{\sqrt{N}} \sum_{i=1}^N |i\rangle_C 
\end{align}
where we assume $|i\rangle_C$ are orthogonal states, and the interaction Hamiltonian takes on the modified form, 
\begin{align}
    \hat{H}_\text{int.} &= \sum_{i=1}^N \hat{\mathcal{H}}_i \otimes |i\rangle\langle i |_C 
\end{align}
where
\begin{align}\label{Hi1det}
    \hat{\mathcal{H}}_i &= \lambda \eta_i(\tau) \sigma_i(\tau) \hat{\Phi}(\mathsf{x}_i(\tau) ) 
\end{align}
contains field operators evaluated along the $i$th trajectory of the superposition. The evolution of the detector-field-control system can be obtained by expanding the time-evolution operator, $\hat{U}$, 
\begin{align}
    \hat{U} &= \sum_{i=1}^N \hat{U}_i \otimes |i\rangle\langle i |_C 
\end{align} 
to second-order in perturbation theory, where
\begin{align}\label{trajectoryunitary}
    \hat{U}_i &= 1 - i\int\D \tau \:\hat{\mathcal{H}}_i - \iint_\mathcal{T} \D \tau \:\D \tau'\hat{\mathcal{H}}_i(\tau ) \hat{\mathcal{H}}_i(\tau') + \mathcal{O}(\lambda^3)
\end{align}
where the second-order terms in $\lambda$ contain time-ordered integrals, denoted by $\mathcal{T}$. Acting $\hat{U}$ on the initial state and measuring the control in the superposition state $|\chi\rangle$ yields the final detector-field state, 
\begin{align}
    |\Psi\rangle_{FD} &= \frac{1}{N} \sum_{i=1}^N \hat{U}_i |0 \rangle |g\rangle .
\end{align}
Upon tracing out the field degrees of freedom, it can be straightforwardly shown that the density matrix of the detector system is given by
\begin{align}
    \hat{\rho}_D &= \begin{pmatrix} 1 - \mathcal{P}_D & 0 \\ 0 & \mathcal{P}_D \end{pmatrix} + \mathcal{O}(\lambda^4)
\end{align}
to leading order in $\lambda$, where 
\begin{align}\label{probability}
    \mathcal{P}_D &= \frac{\lambda^2}{N^2} \sum_{i,j=1}^N \mathcal{P}_{ij,D} = \frac{\lambda^2}{N^2} \left\{ \sum_{i=j}^N \mathcal{P}_{ij,D} + \sum_{i\neq j} \mathcal{P}_{ij,D} \right\} 
\end{align}
is the transition probability of the detector, conditioned on the measurement of the control in the state $|\chi\rangle$. The individual contributions in Eq.\ (\ref{probability}) take the form
\begin{align}\label{transitionprob}
    \mathcal{P}_{ij,D} &= \iint\D \tau \:\D \tau'\eta_{D_j}(\tau ) \eta_{D_i}(\tau') e^{-i\Omega(\tau - \tau') } \mathcal{W}^{ji}\big( \mathsf{x} , \mathsf{x}' \big).
\end{align}
We have utilised the shorthand notation to denote the positive-frequency Wightman functions, 
\begin{align}
    \mathcal{W}^{ji}(\mathsf{x} , \mathsf{x}') &= \langle \psi | \hat{\Phi}\big(\mathsf{x}_i(\tau) \big) \hat{\Phi} \big(\mathsf{x}_j(\tau') \big) | \psi \rangle 
\end{align}
which are two-point field correlation functions evaluated between the trajectories $\mathsf{x}_i(\tau)$ and $\mathsf{x}_j(\tau')$. Equation (\ref{probability}) contains local terms evaluated along the individual trajectories of the superposition, as well as cross-correlation interference terms, evaluated between different trajectories. Notably, these interference terms would be inaccessible to detectors travelling on classical, localised worldlines. If, instead of measuring the control in a superposition basis, one traces out the control state, the final detector state is a probabilistic mixture of the $N$ individual contributions along each trajectory in the superposition, without the interference terms between them:
\begin{align}
    \mathcal{P}_D &= \frac{1}{N} \sum_{i=1}^N \mathcal{P}_{ii,D} .
\end{align}
As we show in Sec.\ \ref{sec:desitter}, the interference terms suppress the local excitations experienced by the individual detectors, which enhances their capacity to extract entanglement from the vacuum state.

\section{Entanglement harvesting with quantum-controlled detectors}\label{sec:eh}

We now consider a composite system of two point-like UdW detectors, coupled to the real, massless scalar field along a quantum-controlled superposition of trajectories. The interaction Hamiltonian is now given by 
\begin{align}
    \hat{H}_\text{int.} &= \sum_{D=A,B} \sum_{i=1}^N \hat{\mathcal{H}}_{D_i} \otimes |i \rangle \langle i|_C
\end{align}
where as in Eq.~\eqref{Hi1det},
\begin{align}
    \hat{\mathcal{H}}_{D_i} &= \lambda \eta_{D_i }(\tau) \sigma_D(\tau)  \hat{\Phi}(\mathsf{x}_{D_i} (\tau) )
\end{align}
describes the part of the interaction between the $D$th detector along the $i$th trajectory of the superposition. To second-order in perturbation theory, the time-evolution operator is
\begin{align}
    \hat{U} &= \sum_{i=1}^N \hat{U}_i \otimes |i\rangle\langle i|_C
\end{align}
where
\begin{align}
    \hat{U}_i &= \mathds{1} + \hat{U}^{(1)} + \hat{U}^{(2)} + \mathcal{O}(\lambda^3)
\end{align}
and $\hat{U}^{(k)}$ is the $k$th order term of the Dyson series expansion, 
\begin{align}\label{18m}
    \hat{U}^{(1)} &= - i\int\D \tau \:\big( \hat{\mathcal{H}}_{A_i}(\tau )  + \hat{\mathcal{H}}_{B_i}(\tau ) \big) \\
    \hat{U}^{(2)} &= - \iint_\mathcal{T} \D \tau \:\D \tau' \big( \hat{\mathcal{H}}_{A_i}(\tau )  + \hat{\mathcal{H}}_{B_i} ( \tau ) \big) \big( \hat{\mathcal{H}}_{A_i} ( \tau' ) + \hat{\mathcal{H}}_{B_i}(\tau' ) \big). \label{18m2}
\end{align}
The initial state of the system is simply
\begin{align}
    |\Psi\rangle_{CFD} &= |\chi\rangle |\psi\rangle |g\rangle_A |g\rangle_B.
\end{align}
Again, let us assume that the control system is measured in its initial state, $|\chi\rangle$. The conditional state of the system after measuring the control in the state $|\chi\rangle$ is
\begin{widetext}
\begin{align}
    \langle \chi | \hat{U} |\Psi\rangle_{CFD} &= \frac{1}{N} \sum_{i=1}^N \Big\{ | \psi\rangle | g\rangle_A | g\rangle_B  + \langle \chi | \hat{U}^{(1)} | \Psi\rangle_{CFD} + \langle \chi | \hat{U}^{(2)} | \Psi\rangle_{CFD} \Big\}
\end{align}
where
\begin{align}\label{25l}
    \langle \chi |\hat{U}^{(1)} | \Psi\rangle_{CFD} &= - i\lambda  \int\D \tau \:e^{i\Omega\tau} \Big( \eta_{A_i } \hat{\Phi}\big( \mathsf{x}_{A_i} ( \tau ) \big) | e\rangle_A |g\rangle_B + \eta_{B_i} \hat{\Phi}\big( \mathsf{x}_{B_i} ( \tau ) \big) |g\rangle_A |e \rangle_B \Big) | \psi\rangle 
\\ \label{26m}
    \langle \chi | \hat{U}^{(2)} | \Psi\rangle_{CFD} &= - \lambda^2 \iint_\mathcal{T} \D \tau \D \tau' e^{-i\Omega(\tau - \tau' ) }  \Big( \eta_{A_i }(\tau ) \eta_{A_i} (\tau' ) \hat{\Phi} \big( \mathsf{x}_{A_i} ( \tau) \big) \hat{\Phi}\big( \mathsf{x}_{A_i} ( \tau' ) \big) + (A \Leftrightarrow B ) \Big)|\psi\rangle | g \rangle_A |g \rangle_B   \nonumber \\
    & - \lambda^2 \iint_\mathcal{T} \D \tau \D \tau' e^{i\Omega(\tau + \tau' ) }  \Big( \eta_{A_i }(\tau ) \eta_{B_i} (\tau' ) \hat{\Phi} \big( \mathsf{x}_{A_i} ( \tau) \big) \hat{\Phi}\big( \mathsf{x}_{B_i} ( \tau' ) \big) + (A \Leftrightarrow B ) \Big)  |\psi\rangle |e \rangle_A |e \rangle_B .
\end{align}
yielding
\begin{align}\label{reduceddensity}
    \hat{\rho}_D &= \begin{pmatrix} 1 - \mathcal{P}_A -\mathcal{P}_B & 0 & 0 & \mathcal{M} \\ 0 & \mathcal{P}_B & \mathcal{L} & 0 \\ 0 & \mathcal{L}^\star  & \mathcal{P}_A & 0 \\ \mathcal{M}^\star & 0 & 0 & 0  \end{pmatrix}
\end{align}
upon tracing out the field degrees of freedom. Note that the $|g\rangle\langle e|_A \otimes |g\rangle \langle e|_B$ element of the density matrix, $\mathcal{M}$,  is obtained from the purely second-order term in the Dyson series expansion (\ref{18m2}) and only contains correlations between the $i$th trajectories of the two detectors.  The 
$|e\rangle\langle g|_A \otimes |g\rangle\langle e|_B$ and $|g\rangle\langle e|_A \otimes |e\rangle\langle g|_B$ terms
in $\hat{\rho}_D$ 
($\mathcal{L}$ and its conjugate) are obtained via a product of the first-order terms in (\ref{18m})
and contain cross-correlations between the $i$th and $j$th trajectories of both detectors. Specifically
\begin{align}
    \mathcal{P}_D &= \frac{\lambda^2}{N^2} \sum_{i,j=1}^N \iint\D \tau \:\D \tau' \big(\eta_{D_i}(\tau ) \eta_{D_j}(\tau' ) e^{-i\Omega(\tau - \tau' ) } \mathcal{W} \big(\mathsf{x}_{D_i}(\tau) ,\mathsf{x}_{D_j}(\tau' ) \big)  \\ 
    \mathcal{L} &= \frac{\lambda^2}{N^2} \sum_{i,j=1}^N \iint\D \tau \:\D\tau' \big( \eta_{B_i }(\tau)  \eta_{A_j}(\tau' ) e^{-i\Omega( \tau - \tau' ) } \mathcal{W} \big(\mathsf{x}_{A_j}(\tau), \mathsf{x}_{B_i}(\tau') \big) \\ 
    \mathcal{M} &= - \frac{\lambda^2}{N} \sum_{i=1}^N \iint_\mathcal{T} \D \tau \: \D \tau' e^{-i\Omega(\tau + \tau' ) } \Big( \eta_{A_i }(\tau )  \eta_{B_i}(\tau' )  \mathcal{W} \big(\mathsf{x}_{A_i}(\tau ) , \mathsf{x}_{B_i}(\tau' ) \big) + \eta_{B_i } (\tau) \eta_{A_i}(\tau' ) \mathcal{W} \big(\mathsf{x}_{B_i } (\tau ) , \mathsf{x}_{A_i}(\tau' ) \big) \Big) .
\end{align}
\end{widetext}
As before, the terms $\mathcal{P}_D$ are local transition probabilities for the $D$th detector. The terminology of `local' in this context must be understood as pertaining to the transition probability of an \textit{individual detector}, recalling that these likewise contain terms evaluated `locally' (in a spatiotemporal sense) along each trajectory of the superposition, as well as `non-locally' between the trajectories (i.e.\ the interference terms). The so-called entangling term, $\mathcal{M}$, is a function of the trajectories of \textit{both} detectors. In particular, it is a sum of the non-local field correlations between the $i$th trajectories of detector $A$ and $B$.  Finally, it should be noted that in our model for the control, the entangling term is identical to that in which the detectors traverse classical worldlines. This is true when the separation of the $i$th superposed trajectories between detectors $A$ and $B$ are equal to the comparable separation of detectors $A$ and $B$ when they traverse classical paths. 

To quantify the entanglement of the final bipartite detector state, we utilise the concurrence, which is defined as \cite{woottersconcurrence}
\begin{align}\label{concurrence}
    \mathcal{C}(\hat{\rho}_D ) &= 2 \text{max} \left\{ 0 , |\mathcal{M} | - \sqrt{\mathcal{P}_A\mathcal{P}_B} \right\}.
\end{align}
From Eq.\ (\ref{concurrence}), we see that the amount of entanglement extracted from the field is a competition between the entangling term, $\mathcal{M}$, and the geometric mean of the local transition probabilities of the two detectors, $\sqrt{\mathcal{P}_A\mathcal{P}_B}$. For entanglement to be extracted from the field, the non-local correlations probed by the detectors must exceed the local field fluctuations perceived along their respective (superposed) trajectories. In this paper, we consider detectors with equal transition probabilities (i.e.\ in de Sitter, experiencing equal redshifts) so that the concurrence simplifies to 
\begin{align}
    \mathcal{C} ( \hat{\rho}_D ) &= 2 \text{max} \left\{ 0 , | \mathcal{M} | - \mathcal{P}_D \right\} 
\end{align}
where $\mathcal{P}_D$ is just the transition probability of either of the detectors. 

Another quantity of interest is the \textit{mutual information} harvested by the detectors, which quantifies the total amount of correlations -- both quantum and classical -- between them. For bipartite quantum systems, the mutual information is given by the relative entropy between the state of the system and the tensor product of the reduced states of the subsystems,
\begin{align}
    I(\rho_{AB}) &= S \big( \rho_{AB} || \rho_{A } \otimes \rho_B \big) .
\end{align}
From the reduced density matrix obtained in Eq.\ \eqref{reduceddensity}, the mutual information between two UdW detectors to leading order in $\lambda$, is \cite{simidzijathermalstates}
\begin{align}\label{mutinf}
    I(\rho_{AB}) &= \mathcal{L}_+ \log \mathcal{L}_+ + \mathcal{L}_- \log \mathcal{L}_- - \mathcal{P}_A \log P_A - \mathcal{P}_B \log \mathcal{P}_B
\end{align}
where
\begin{align}\label{lpm}
    \mathcal{L}_\pm &= \frac{1}{2} \Bigg\{ \mathcal{P}_A + \mathcal{P}_B \pm \sqrt{\big( \mathcal{P}_A - \mathcal{P}_B\big)^2 + 4 \big| \mathcal{L} \big|^2 }\Bigg\} .
\end{align}
Since we consider detectors with equal transition probabilities, Eq.\ (\ref{lpm}) simplifies to 
\begin{align}
    \mathcal{L}_\pm &= \mathcal{P}_D \pm | \mathcal{L}| .
\end{align}
For detectors which have vanishing concurrence but non-zero mutual information, we can infer that the correlations are either classical, or come from non-entanglement quantum correlations known as quantum discord. Note in particular that the mutual information contains the off-diagonal $\mathcal{L}$ terms of the detector density matrix. Interestingly, $\mathcal{L}$ contains Wightman functions evaluated along the $j$th trajectory of detector $A$, and the $i$th trajectory of detector $B$, which include correlations that are not directly probed by the detectors, upon inclusion of the quantum-controlled superposition. These terms can be understood as the leading order contributions to the classical correlations between the detectors \cite{sachspairsquadratic}. 

\section{Entanglement Harvesting in de Sitter Spacetime}\label{sec:desitter}
\subsection{Field-theoretic and geometric details}
de Sitter spacetime is a maximally symmetric spacetime of constant positive curvature that is a solution to the Einstein field equations with cosmological constant $\Lambda$. We shall write 
the  inverse de Sitter length
$l = \sqrt{\Lambda/3} $
  \cite{barrabes2008einstein,griffiths2009exact,salton2015acceleration,nambu2013entanglement}, as it allows us to refer to $l$ interchangeably with the spacetime curvature, since it is related to the Ricci scalar via $R = 12l^2$. 
  
  The de Sitter manifold can be described by the 5-dimensional hyperboloid, \begin{align}
    -Z_0^2 + Z_1^2 + Z_2^2 + Z_3^2 + Z_4^2 &= 1/l^2. 
\end{align}
 The hyperboloid is embedded within a flat 5-dimensional geometry, 
\begin{align}
    \D s^2 &= - \D Z_0^2 + \D Z_1^2 + \D Z_2^2 + \D Z_3^2 + \D Z_4^2
\end{align}
with coordinates $(Z_0, Z_1, Z_2, Z_3, Z_4)$. A particularly convenient coordinate system useful for field-theoretic calculations are the static coordinates $(t,r,\theta,\phi)$, given by 
\begin{align}\label{coordinates}
    Z_0 &= \sqrt{1/l^2 - r^2}\sinh (lt) \\
    Z_1 &= \sqrt{1/l^2 - r^2} \cosh(lt) \\
    Z_2 &= r \cos\theta \vphantom{\sqrt{1/l^2 - r^2}} \\
    Z_3 &= r \sin\theta\cos\phi  \vphantom{\sqrt{1/l^2 - r^2}} \\ \label{Z4}
    Z_4 &= r \sin\theta \sin\phi  \vphantom{\sqrt{1/l^2 - r^2}}
\end{align}
with the corresponding metric given by 
\begin{align}
    \D s^2 &= - f(r) \D t^2 + \frac{\D r^2}{f(r)}+ r^2 \D \Omega_2^2
\end{align}
where
\begin{align}
    f(r) &= 1 - l^2r^2
\end{align}
and $\D \Omega_2^2 = \D \theta^2 + \sin^2\theta\D \phi^2$. The coordinates Eq.\ (\ref{coordinates})--(\ref{Z4}) only cover part of the entire de Sitter spacetime, a region known as the static patch (for a useful visualisation, see \cite{griffiths2009exact}). A test particle in this spacetime experiences an acceleration with magnitude, 
\begin{align}
    \alpha &= \frac{lR_D}{\sqrt{l^{-2} - R_D^2}},
\end{align}
where $r = R_D$ is the radial coordinate of the particle. Evidently, both the curvature of the spacetime and the radial position of the test particle influence its acceleration. Furthermore, as $l\to 0$ or $R_D\to 0$, the acceleration vanishes. 

We consider a massless, conformally coupled real scalar field since the Wightman function takes on the simple form \cite{birrell1984quantum,polarski1989hawking},
\begin{align}
    \mathcal{W}_\text{dS} &= - \frac{1}{4\pi^2} \frac{1}{(Z_0 - Z_0')^2 - \Delta Z_i^2 - i\varepsilon}
\end{align}
where $\Delta Z_i^2 = (Z_1 - Z_1')^2 + (Z_2 - Z_2')^2 + (Z_3 - Z_3')^2 + (Z_4-Z_4')^2$ and $\varepsilon$ is an infinitesimal regularisation constant. We consider detectors superposed along the trajectories
\begin{align}
    &\:\:\:\underbrace{(R_D,\theta_1,\phi),  (R_D,\theta_2,\phi)}_\text{detector $A$}  \\
    &\underbrace{(R_D,\theta_1',\phi), (R_D,\theta_2',\phi)}_\text{detector $B$}
\end{align}
respectively (i.e.\ a two-trajectory superposition, for simplicity), such that the $i$th set of trajectories are separated by the Euclidean (entangling) distance \cite{tian2016detecting}
\begin{align}
    \mathcal{L}_{Mi} &= 2 R_D \sin \left( \frac{\theta_{Mi}}{2} \right)
\end{align}
where $ \theta_{Mi} = \theta_i' - \theta_i  > 0$. In our analysis, we take $\theta_M := \theta_{M1} = \theta_{M2}$, and assume that both detectors are at equal $r$, $\phi$, for simplicity. This also means that the entangling term of the bipartite detector density matrix is identical to the classical trajectory case. We make use of three kinds of Wightman function: $\mathcal{W}_\text{dS}$, a self-correlation term evaluated locally along each trajectory in the superposition, for each detector, $\mathcal{W}_\text{dS-$s$}$, evaluated non-locally between the superposed trajectories of a single detector, which acts as an interference term between the trajectories, and $\mathcal{W}_\text{dS-$m$}$, evaluated between the $i$th trajectories of the respective detectors, $A$ and $B$, and carries information about the non-local field correlations probed by the two detectors. 

Explicitly, the Wightman functions are given by 
\begin{align}\label{wightmansingle}
    \mathcal{W}_\text{dS}(s) &= - \frac{\beta^2}{16\pi^2} \frac{1}{\sinh^2(\beta s/2 - i\varepsilon)} \\ \label{wightmandouble}
    \mathcal{W}_\text{dS-$s$}(s) &= - \frac{\beta^2}{16\pi^2} \frac{1}{\sinh^2(\beta s/2 - i\varepsilon) - (\beta \mathcal{L}_S/2)^2} \\
    \mathcal{W}_\text{dS-$m$}(s) &= - \frac{\beta^2}{16\pi^2} \frac{1}{\sinh^2(\beta s/2 - i\varepsilon) - (\beta \mathcal{L}_M/2)^2} 
\end{align}
where $\beta = ( l^{-2} - R_D^2 )^{-1/2}$. Now, let us assume for illustration that a single detector interacts with the field mediated by a Gaussian switching function in the infinite interaction-time limit (that is, the detector-field interaction is effectively constant). It can be straightforwardly shown that the transition probability of a single detector in this limit is proportional to the Planck spectrum \cite{gibbons1977cosmological},
\begin{align}
    \mathcal{P}_D &= \frac{\lambda^2\Omega}{2\pi} \frac{1}{e^{2\pi\Omega/\beta} - 1},
\end{align}
meaning that the field is populated by thermal particles at the Gibbons-Hawking temperature
\begin{align}
    T_\text{dS} &= \frac{\beta}{2\pi}. 
\end{align}
In Eq.\ (\ref{wightmandouble}), we have also defined the superposition distance, $\mathcal{L}_S$, as
\begin{align}
    \mathcal{L}_S &= 2R_D \sin\left( \frac{\theta_S}{2} \right)
\end{align}
which is the Euclidean distance separating the $i$th and $j$th trajectories of the superposition, where $\theta_S = |\theta_i - \theta_j|$ is the angular separation between them. Furthermore, since we have assumed $\theta_{M1} = \theta_{M2}$, it follows that the superposition distances (i.e.\ dependent on the angular separations, $\theta_S$ and $\theta_S'$, over which they are superposed) of both detectors are also equal.

The simple form of the Wightman functions allows for the direct numerical evaluation of the transition probability and the entangling term in the reduced density matrix. In the following analysis, we consider entanglement harvesting between detectors in spatial superpositions, as well scenarios where the detector-field interaction occurs in a superposition of temporal order. We compare our results with scenarios without the superposition, demonstrating in general, that superposition enhances the amount of harvested entanglement.

\subsection{Spatial superpositions}
\begin{figure}[h]
    \centering
    \includegraphics[width=\linewidth]{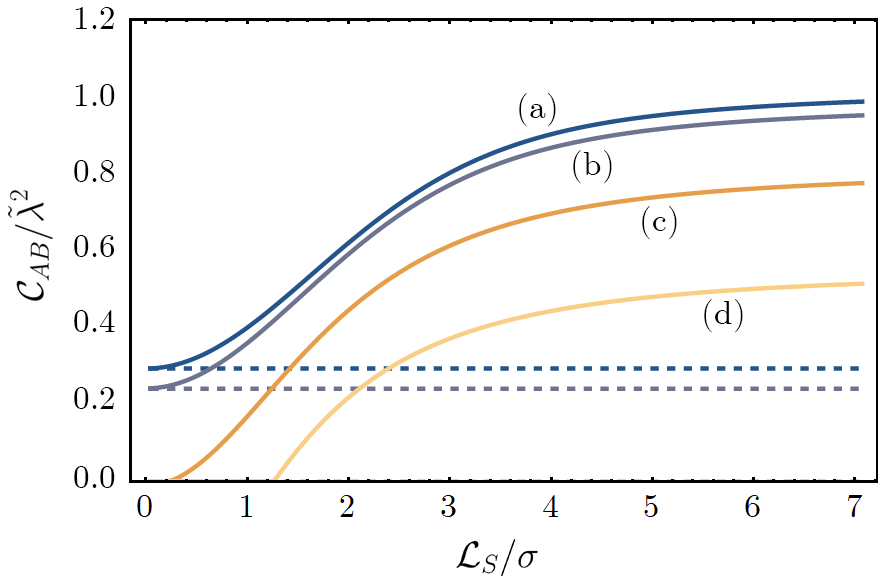}
    \caption{Concurrence, $\mathcal{C}_{AB}/\tilde{\lambda}^2$ as a function of the superposition distance, $\mathcal{L}_S$ for (a) $l\sigma = 0.1$, (b) $l\sigma = 0.175$ (c) $l\sigma = 0.195$ and (d) $l\sigma = 0.198$. The dashed lines represent the concurrence harvested by detectors on classical trajectories corresponding to cases (a), (b); for (c) and (d), it is impossible for such detectors to harvest entanglement. The other parameters we have used are $R_D/\sigma = 5$, $\Omega\sigma = 0.05$, $\mathcal{L}_M /\sigma = 1.36$. }
    \label{fig:deistterspatialLS}
\end{figure}

We  consider first detectors in a superposition of two trajectories, where the initial state of the control is given by 
\begin{align}
    |\chi\rangle &= \frac{1}{\sqrt{2}} \big( |1\rangle_C + |2 \rangle_C \big) .
\end{align}
Likewise, we assume that the final measurement of the control in the state $|\chi\rangle$. We take the switching function to be identical along each trajectory, namely a Gaussian centred at $\tau = 0$, 
\begin{align}
    \eta_{D_i}(\tau) &= \exp \left\{ - \frac{\tau^2}{2\sigma^2} \right\}
\end{align}
where $\sigma$ is a characteristic timescale for the interaction. The excitation probability for both detectors is given by 
\begin{align}
    \mathcal{P}_D &= \frac{\lambda^2 \sqrt{\pi \sigma^2}}{2} \Bigg\{  \infint\D s \: \frac{e^{-s^2/4\sigma^2}e^{-i\Omega s}}{\sinh^2(\beta s/2 -i \varepsilon)} \nonumber  \\
    & + \infint\D s \:\frac{e^{-s^2/4\sigma^2} e^{-i\Omega s}}{\sinh^2(\beta s/2-i\varepsilon) - (\beta\mathcal{L}_S/2)^2} \Bigg\} 
\intertext{where the contribution from `local' excitations (i.e.\ along the individual trajectories) and `non-local' interference terms is evident. Meanwhile, the entangling term takes the form}
    \mathcal{M} &= - 2\lambda^2 \sqrt{\pi \sigma^2} e^{-\sigma^2\Omega^2} \int_0^\infty \frac{\D s\:e^{-s^2/4\sigma^2}}{\sinh^2(\beta s/2-i\varepsilon) - ( \beta\mathcal{L}_M/2)^2 }.
\end{align}
In Fig.\ \ref{fig:deistterspatialLS}, we have plotted the concurrence between the detectors as a function of the superposition distance, $\mathcal{L}_S /\sigma$. 
For the detectors in superposition, the concurrence is a monotonically increasing function of the superposition distance, asymptoting to a fixed value at large distances. Since the entangling term is identical for both vanishing and non-vanishing superposition distances, we can trace this behaviour to the interference terms in the local transition probabilities for the individual detectors. For small separations between the superposed trajectories, these terms are approximately equal to the contributions evaluated locally along the individual trajectories. As the separation increases, the interference terms decay rapidly, a generic feature of non-local field correlations. This suppresses the excitations experienced by the individual detectors, allowing the entangling term, $\mathcal{M}$, to dominate.  Such an effect was first demonstrated  for temporally superposed detectors in Minkowski spacetime   \cite{henderson2020quantum}.

\begin{figure}[h]
    \centering
    \includegraphics[width=\linewidth]{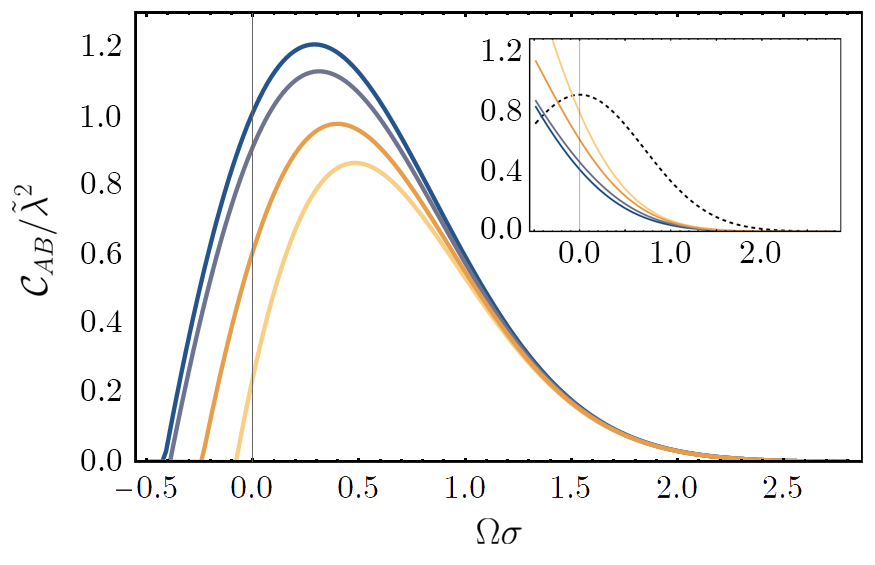}
    \caption{Concurrence, $\mathcal{C}_{AB}/\tilde{\lambda}^2$ as a function of the energy gap, $\Omega\sigma$, comparing the entanglement harvested by detectors on classical trajectories, with those in superposition. The plot displays the concurrence for detectors with $\theta_S = 0$, $\pi/8$, $\pi/4$, $\pi/2$ (from yellow to dark blue). The inset compares the entangling term, $\mathcal{M}$ (dashed line), with the transition probability $\mathcal{P}_D$. The other parameters used are $a \sigma = 0.15, R_D/\sigma = 5$, and $\theta_M = \pi/12$.}
    \label{fig:spatialdesittervsW}
\end{figure}

In Fig.\ \ref{fig:spatialdesittervsW}, we have plotted the concurrence as a function of the detector energy gap, $\Omega\sigma$. In general, the concurrence rapidly decays for negative energy gaps (i.e.\ detectors initialised in their excited state). As before, the presence of the superposition amplifies the concurrence, which again, can be understood in terms of a suppression effect in the local noise term, $\mathcal{P}_D$, for larger superposition distances (see inset). At large energy gap, this term becomes independent of the superposition distance. 

\begin{figure}[h]
    \centering
    \includegraphics[width=\linewidth]{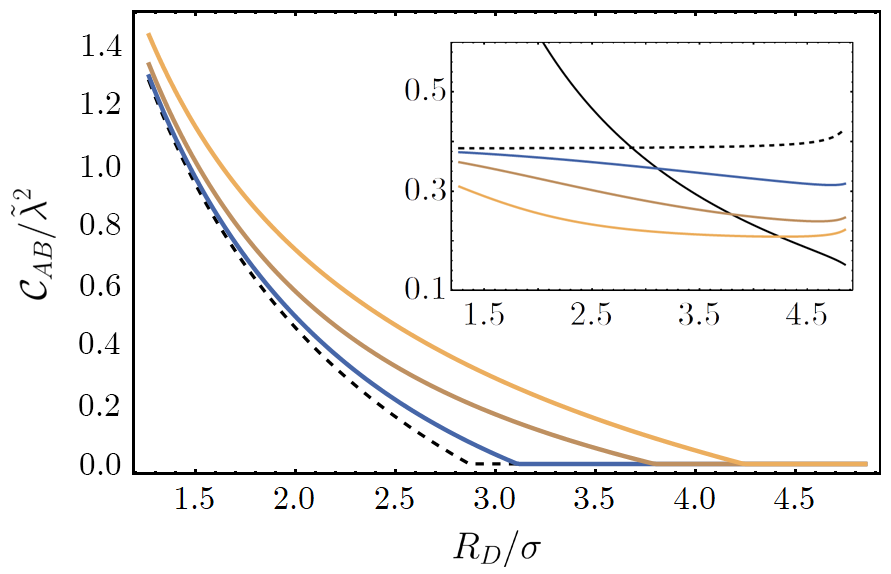}
    \caption{Concurrence, $\mathcal{C}_{AB}/\tilde{\lambda}^2$ as a function of the radial coordinate, $R_D/\sigma$ where we have compared detectors on classical trajectories with detectors in superposition, with (blue) $\theta_S = \pi/8$, (brown) $\theta_S = \pi/4$, and (orange) $\theta_S = \pi/2$. The dashed line in the main figure is the concurrence for detectors on classical trajectories. The other parameters used are $l\sigma = 0.2$, and $\Omega\sigma = 0.02$. The inset compares the entangling term, $\mathcal{M}$ (black line), with the local noise term, $\mathcal{P}_D$, where the colours correspond to those shown in the main figure.}
    \label{fig:desitterspatialVSR}
\end{figure}

In Fig.\ \ref{fig:desitterspatialVSR}, we have plotted the concurrence as a function of the radial coordinate, $R_D/\sigma$ for a fixed spacetime curvature, $l\sigma$. As before, the superposition of each detector's path inhibits the local excitations it perceives, and larger separations produce a more significant suppression of these terms. Remarkably, the presence of the interference terms actually causes the transition probability to decay as the detectors approach the horizon. That is to say that as the temperature of the field increases, the excitation probability of the detectors \textit{decreases}, in the regimes shown.  Such a behaviour has been observed in other contexts, and has been referred to as the anti-Unruh effect 
\cite{brenna2016anti,garay2016thermalization} (for flat spacetime)  and the anti-Hawking effect \cite{henderson2019btz} (in the vicinity of a black hole).
This behaviour is explored in greater detail in a companion paper  (in preparation). 
 Here, we find that such an effect enables entanglement harvesting in regimes that are not be possible for detectors on classical paths. 

\subsection{Temporal superpositions}
Thus far, we have considered detectors travelling on spatially delocalised trajectories. In particular, each detector was initialised in a superposition of physical paths, separated by a fixed Euclidean distance. We now turn to detectors travelling along single trajectories whose interaction with the field occurs in a \textit{temporal} superposition. 

In the case of temporal superposition, the temporal switching of each detector occurs in a quantum-controlled superposition. As before, the control state is initialised and then measured in the superposition state
\begin{align}
    |\chi_+ \rangle &= \frac{1}{\sqrt{2}} \big( |1 \rangle + |2 \rangle \big) .
\end{align}
In the following, we refer to such detectors as interacting with `non-classical' spacetime regions (i.e.\ having switching functions subjected to quantum indeterminacy). There are two scenarios of particular interest. The first is what was labelled in \cite{henderson2020quantum} as a \textit{past-future} superposition, in which the detectors are jointly switched on along a common spacelike slice, but in a quantum-controlled superposition of times (i.e.\ the centre-time of the Gaussian). The switching functions of the two detectors take the form, 
\begin{align}
    \eta_{A_1}(\tau) = \eta_{B_1}(\tau) &= \exp \bigg\{ - \frac{(\tau + \tau_0)^2 }{2\sigma^2} \bigg\} \\
    \eta_{A_2}(\tau) = \eta_{B_2}(\tau) &= \exp \bigg\{ - \frac{(\tau - \tau_0 )^2}{2\sigma^2} \bigg\}.
\end{align}
Here $-\tau_0$ (which we refer to as the superposition time-delay, defined with respect to the detector frame) is the time-delay of the switching function with respect to the origin. The second scenario, labelled \textit{cause-effect} superpostion \cite{henderson2020quantum}, is one in which the causal order of the interactions of the two detectors with the field is indefinite. This is achieved by a superposition of the causal arrangements of the two detectors; that is, detector $A$ switched on in the causal past of detector $B$, in superposition with detector $B$ switched on in the causal past of detector $A$. The switching functions take the form,
\begin{align}
    \eta_{A_1}(\tau) = \eta_{B_2}(\tau) &= \exp \bigg\{ - \frac{(\tau + \tau_0)^2}{2\sigma^2} \bigg\} \\
    \eta_{A_2}(\tau) = \eta_{B_1}(\tau) &= \exp \bigg\{ - \frac{(\tau - \tau_0)^2}{2\sigma^2} \bigg\}.
\end{align}
Both the past-future and cause-effect superpositions represent scenarios possessing \textit{indefinite causal order}. In \cite{henderson2020quantum}, it was shown that detectors interacting with the Minkowski vacuum in a quantum-controlled superposition of temporal orders (i.e.\ centre switching times) can harvest entanglement under generic conditions, violating the no-go theorem of \cite{simidzija2018general} for detectors activated only once. The setup in \cite{henderson2020quantum} also represents a toy model for the quantum switch \cite{chiribella2019quantum}, a scenario where operations on a quantum system occur in a superposition of time orders, which was first proposed within the framework of the process matrix formalism. The following analysis represents a next step in understanding the effect of indefinite causal order upon the ability for detectors to harvest entanglement, with the eventual aim to apply it to gravitationally-induced indefinite causal order \cite{zych2019bell}.   

\subsubsection{Past-future superposition}

We first  consider entanglement harvesting with detectors in a past-future superposition. The local transition probability   and entangling terms are respectively given by 
\begin{align}\label{pastfutureprobability}
    \mathcal{P}_D &= \frac{\lambda^2\sqrt{\pi\sigma^2}}{2} \Bigg\{ \infint \D s \: \frac{e^{-i\Omega s} e^{-s^2/4\sigma^2}}{\sinh^2(\beta s/2-i\varepsilon)} \nonumber \\ 
    & + e^{-\tau_0^2/\sigma^2 } \infint\D s \: \frac{e^{-i\Omega s } e^{-s^2/4\sigma^2} \cosh \left( s\tau_0/\sigma^2 \right)}{\sinh^2(\beta s/2-i\varepsilon)} \Bigg\} 
\intertext{and}
    \label{MPFdesitter}
    \mathcal{M} &= - 2 \sqrt{\pi \sigma^2} \cos \left( \frac{\tau_0 \Omega}{2} \right) e^{-\sigma^2\Omega^2} \nonumber \\
    & \times \int_0^\infty\D s\: \frac{e^{-s^2/4\sigma^2}}{\sinh^2(\beta s/2-i\varepsilon) - (\beta\mathcal{L}_M/2)^2}  .
\end{align}
Notably, the entangling term is an oscillatory function of the centre interaction time, $\tau_0$, and the energy gap, $\Omega$. As $\tau_0$ is varied, the entangling term will experience periodic resonances, produced by interference between the field regions that the detector interacts with in temporal superposition. Unlike   cause-effect superposition, the entangling term does not decay with $\tau_0$. This is because the $i$th interaction region of detector $A$ and $B$ remain at a fixed distance in spacetime, independent of $\tau_0$. Finally, for detectors with classical switching functions, the concurrence is independent of $\tau_0$, since the Wightman functions for the cases considered are time-translation invariant.
\begin{figure}[h]
    \centering
    \includegraphics[width=\linewidth]{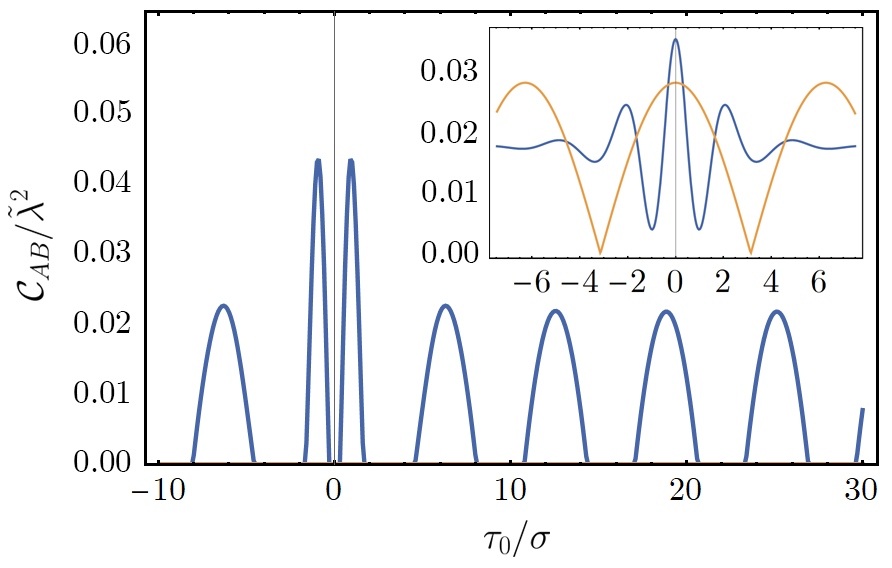}
    \caption{The concurrence $\mathcal{C}_{AB}/\tilde{\lambda}^2$ harvested by detectors in a past-future temporal superposition, as a function of the superposition time-delay $\tau_0/\sigma$ for $l\sigma = 0.2$, $R_D/\sigma = 2.5$, $\Omega\sigma =1$, $\mathcal{L}_M/\sigma = 3.53$. Detectors on classical trajectories in this regime cannot harvest entanglement. The inset compares the entangling term $\mathcal{M}$ (orange curve), with the local noise term, $\mathcal{P}_D$ as a function of $\tau_0/\sigma$. }
    \label{fig:desitterPF1}
\end{figure}

In Fig.\ \ref{fig:desitterPF1}, we have plotted the concurrence harvested by the detectors as a function of the superposition time-delay, $\tau_0/\sigma$. In the regime considered, entanglement harvesting is possible for detectors in superposition, but impossible for detectors with classical switching functions. We notice that the entangling term oscillates periodically as a function of $\tau_0/\sigma$, as was already inferred from the form of Eq.\ (\ref{MPFdesitter}). The transition probability is also oscillatory and decays to an equilibrium value as $\tau_0/\sigma$ grows large. This equilibrium value is half  that of a single detector; only the `local' contributions to $\mathcal{P}_D$ are non-vanishing at large separations. For $\tau_0/\sigma = 0$, there is no superposition of interaction times, and the transition probabilities equal that of a single detector. Correspondingly, we find that entanglement harvesting is generally inhibited for $|\tau_0/\sigma| \sim 0$. Interestingly, for $|\tau_0/\sigma|\ll 1$ (on either side of $|\tau_0/\sigma|\sim 0$), the local field excitations of each detector are significantly inhibited
, which allows for an \textit{enhancement} of the concurrence, indicated by the sharp peaks near the origin in Fig.~\ref{fig:desitterPF1}.
As $|\tau_0/\sigma|$ grows, the temporal distance between the interaction regions likewise grows, and the interference terms in the transition probability decay to zero. Hence, the equilibrium value for the transition probability is simply half of that for a single detector interacting with the field in a localised spacetime region. 

\begin{figure}[h]
    \centering
    \includegraphics[width=0.8\linewidth]{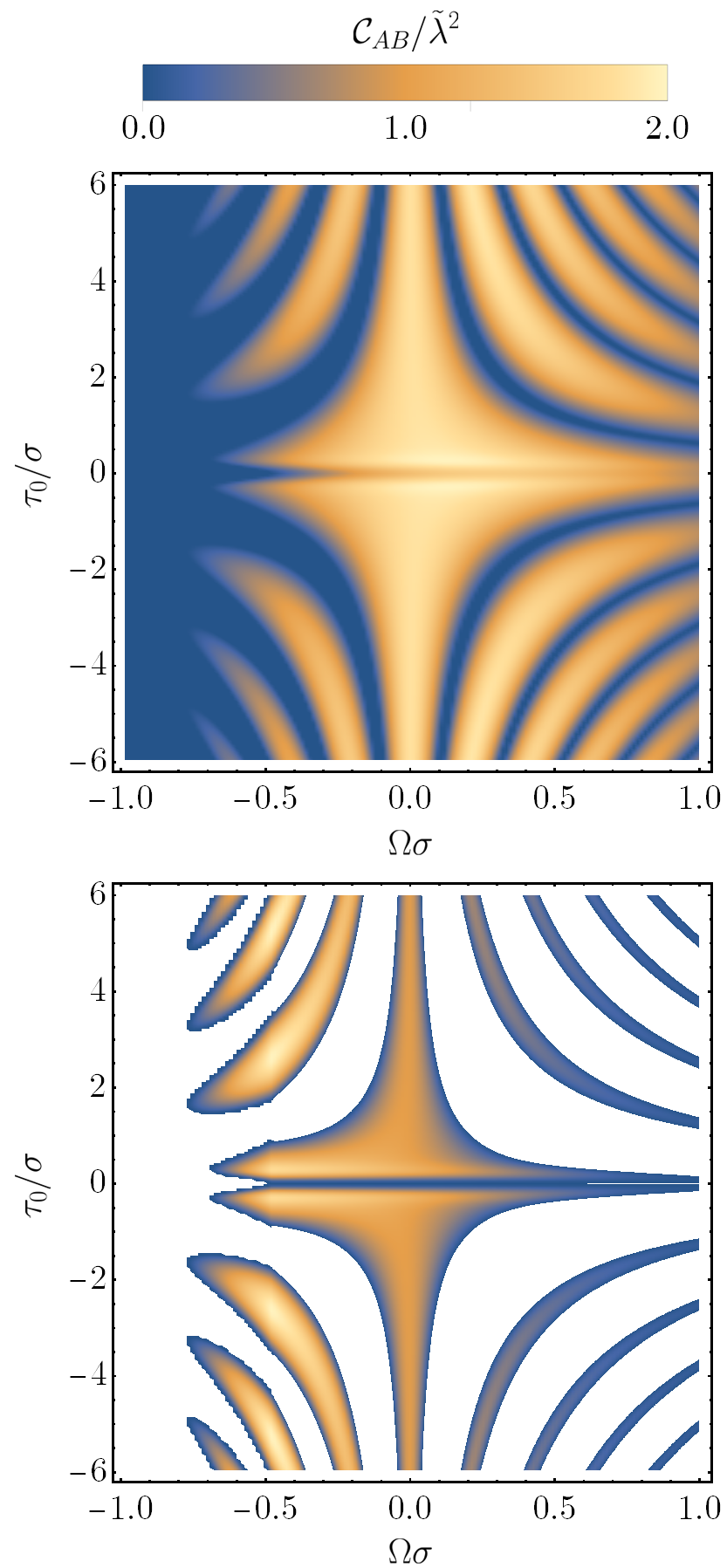}
    \caption{(top) Plot of the concurrence, $\mathcal{C}_{AB}/\tilde{\lambda}^2$, as a function of the past-future superposition time-delay   $\tau_0/\sigma$ and the energy gap, $\Omega\sigma$, with $l\sigma = 0.2$, $R_D/\sigma = 2.5$, $\mathcal{L}_M /\sigma = 0.65$. (bottom) Plot of the difference in the concurrence obtained by the past-future detector superposition of the top figure, compared with an analogous setup without the superposition. The coloured regions are those in which superposed detectors can harvest entanglement where classical detectors cannot, and vice versa for the white regions.}
    \label{fig:past-futuredensity}
\end{figure}

In Fig.\ \ref{fig:past-futuredensity}, we have plotted the concurrence as a function of the energy gap, $\Omega\sigma$, and the superposition time-delay, $\tau_0/\sigma$. In particular, the bottom graph displays the regions of the parameter space for which larger amounts of entanglement are harvested by the quantum-controlled detectors, in comparison to detectors on classical trajectories. Some features of note include the amplification of entanglement harvesting at small negative energy gaps, as well as the regions near $|\tau_0/\sigma| \sim 0$ for which the resonant trough in the transition probability amplifies the concurrence between the detectors. 

\subsubsection{Cause-effect superposition}
\begin{figure}[h]
    \centering
    \includegraphics[width=\linewidth]{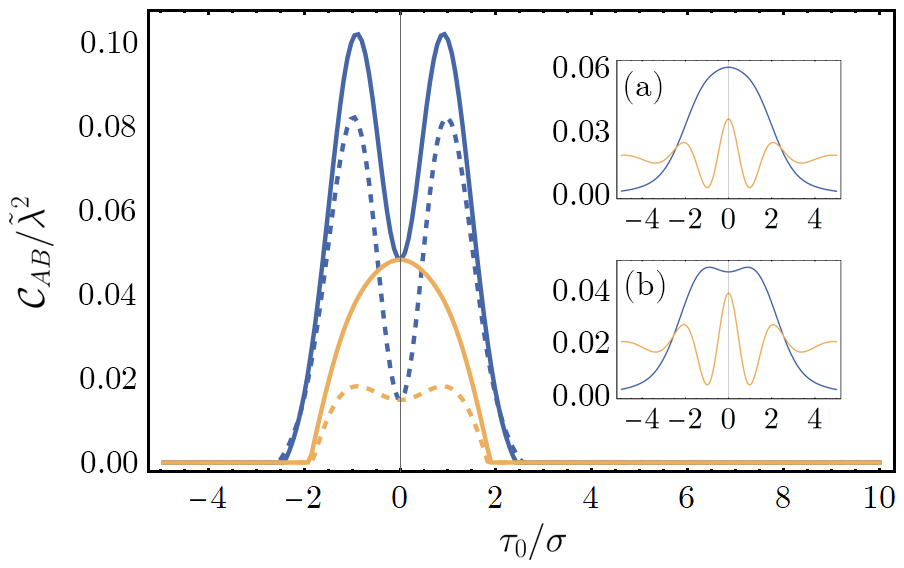}
    \caption{Concurrence, $\mathcal{C}_{AB}$, harvested by detectors with classical switching times (orange) compared to a cause-effect superposition of interaction times (blue), as a function of $\tau_0/\sigma$. The dashed lines correspond to $\mathcal{L}_M/\sigma  = 3.53$, while the solid lines correspond to $\mathcal{L}_M = 2.5$. The insets show the entangling term (blue) and local noise (orange) for (a) $\mathcal{L}_M/\sigma  = 2.5$ and (b) $\mathcal{L}_M/\sigma  = 3.53$. The other parameters we have used are $l\sigma = 0.2$, $R_D/\sigma = 2.5$, $\Omega\sigma = 1$. }
    \label{fig:desitterCE1}
\end{figure}

For the cause-effect superposition, the transition probability of each detector takes the same form as the past-future superposition, Eq.\ (\ref{pastfutureprobability}). The entangling term decays with $\tau_0$, since the spacetime distance between the $i$th interaction region of detector $A$ and $B$ increases with $\tau_0$. The local transition probability and entangling terms are given respectively by 
\begin{align}
    \mathcal{P}_D &= \frac{\lambda^2\sqrt{\pi\sigma^2}}{2} \Bigg\{ \infint \D s \: \frac{e^{-i\Omega s} e^{-s^2/4\sigma^2}}{\sinh^2(\beta s/2-i\varepsilon)}  \nonumber \\ 
    & + e^{-\tau_0^2/\sigma^2 } \infint\D s \: \frac{e^{-i\Omega s } e^{-s^2/4\sigma^2} \cosh \left( s\tau_0/\sigma^2 \right)}{\sinh^2(\beta s/2-i\varepsilon)} \Bigg\} 
    \intertext{and}
    \mathcal{M} &= 4 \sqrt{\pi \sigma^2} e^{-\sigma^2\Omega^2} e^{-\tau_0^2/\sigma^2} \nonumber \\
    & \times \int_0^\infty\D s \: \frac{e^{-s^2/4\sigma^2} \cosh \left( s\tau_0/\sigma^2 \right)}{\sinh^2(\beta s/2-i\varepsilon) - (\beta\mathcal{L}_M/2)^2}
\end{align}
In Fig.\ \ref{fig:desitterCE1}, we have plotted the concurrence as a function of $\tau_0/\sigma$, comparing both the cause-effect arrangement and the scenario where the detectors $A$ and $B$ interact with the field in classical spacetime regions, with a time-delay. 

Unlike the past-future superposition, which amplified entanglement harvesting at periodic values of $\tau_0/\sigma$, the cause-effect superposition enables in general, a greater amount of entanglement to be harvested, compared with detectors with classical switching functions. Furthermore, the cause-effect superposition also allows entanglement to be harvested in regimes where it would not be possible with classical detectors. 

By construction, the interference terms equal the local contributions when $\tau_0/\sigma = 0$, and hence $\mathcal{P}_D$ is maximised. For small superposition distances, $|\tau_0/\sigma|\ll 1$, the transition probability of each detector experiences an anti-resonance (i.e.\ dipping below the $|\tau_0/\sigma|\gg 1$ equilibrium value), amplifying the concurrence between the detectors. For larger time-delays, the entangling term decays rapidly and the detectors can no longer harvest entanglement. 
\begin{figure}[h]
    \centering
    \includegraphics[width=0.825\linewidth]{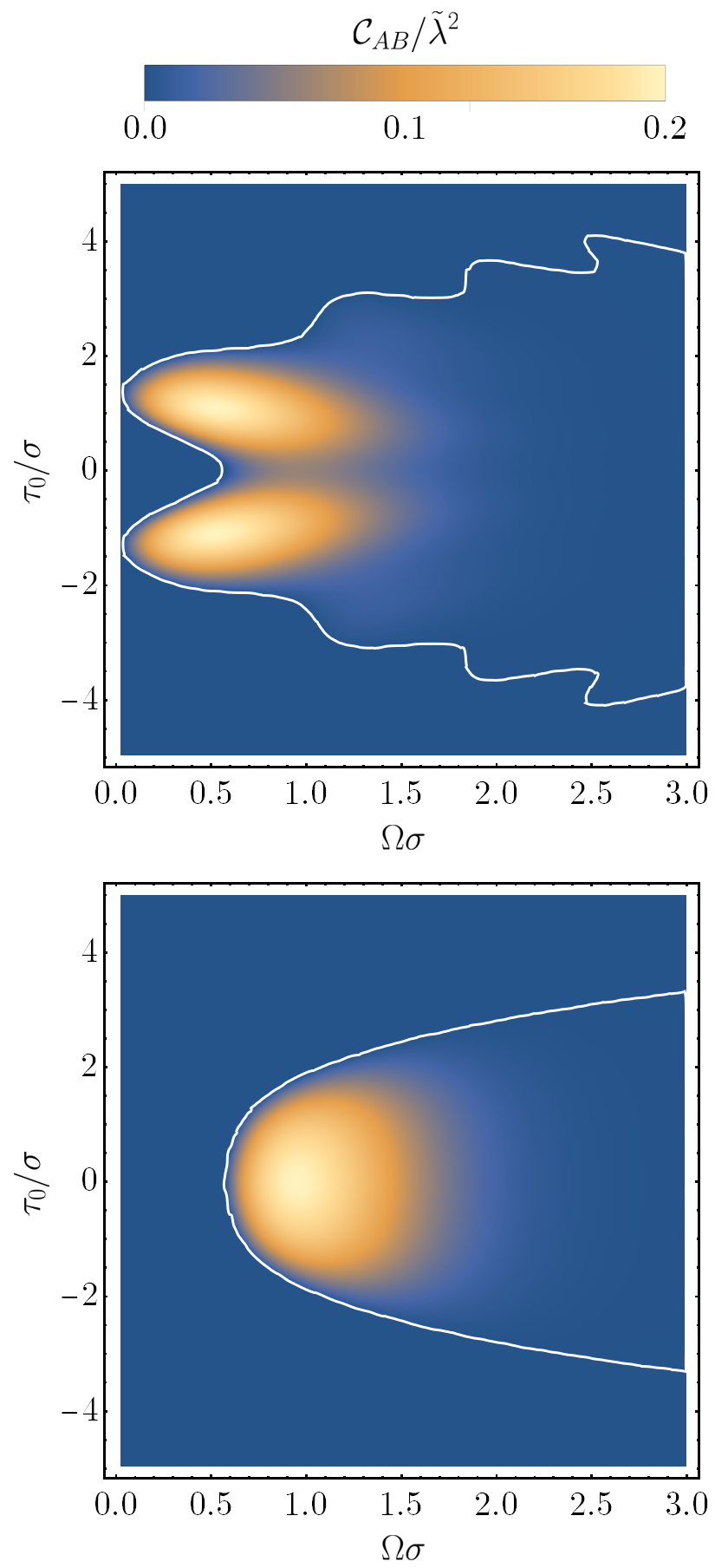}
    \caption{(top) Concurrence, $\mathcal{C}_{AB}/\tilde{\lambda}^2$, as a function of $\tau_0/\sigma$ and $\Omega\sigma$ for detectors in a cause-effect superposition of interaction times. The white line represents the contour of zero concurrence. We have used the parameters $l\sigma = 0.2$, $R_D/\sigma = 2.5$, and $\mathcal{L}_M/\sigma = 2.5$. (bottom) Concurrence between detectors in an analogous setup, without the superposition of interaction times. For all parameter regimes where the cause-effect superposed detectors are entangled, the amount of entanglement is larger than that obtained for detectors with classical switching functions. }
    \label{fig:densitycauseeffect}
\end{figure}

In Fig.\ \ref{fig:densitycauseeffect}, we compare the concurrence harvested by the detectors for (a) detectors in a cause-effect superposition and (b) detectors interacting with the field in classically defined spacetime regions. In general, the detectors in superposition can harvest  more entanglement than the detectors with classically defined interaction regions; the presence of the superposition also yields non-vanishing concurrence at larger values of $\tau_0/\sigma$ than would be possible for the analogous classical-switching setup. Hence, the interference between the different spacetime regions probed by the delocalised detectors allows entanglement harvesting to occur at larger time-delays than would be possible in the corresponding classical case, as evidenced by the oscillations in Fig.\ \ref{fig:densitycauseeffect}(a).

\section{Harvesting mutual information}

\subsection{Spatial superpositions}

So far, we have studied the harvesting of entanglement by detectors in superposition. We now turn to the harvesting of mutual information, Eq.\ \eqref{mutinf}. The local transition probability of each detector is the same as that derived previously for spatial superpositions, while we additionally require the nonlocal correlation term of the detector density matrix, $\mathcal{L}$. These are given respectively by
\begin{widetext}
\begin{align}\label{64}
    \mathcal{P}_D &= \frac{\lambda^2 \sqrt{\pi \sigma^2}}{2} \Bigg\{  \infint\D s \: \frac{e^{-s^2/4\sigma^2}e^{-i\Omega s}}{\sinh^2(\beta s/2-i\varepsilon) } + \infint\D s \: \frac{e^{-s^2/4\sigma^2} e^{-i\Omega s} }{\sinh^2(\beta s/2-i\varepsilon) - (\beta\mathcal{L}_S/2)^2} \Bigg\}  
\\
    \mathcal{L} &= \frac{\lambda^2 \sqrt{\pi\sigma^2}}{2} \Bigg\{ \infint \D s \: \frac{e^{-s^2/4\sigma^2}e^{-i\Omega s}}{\sinh^2(\beta s/2-i\varepsilon) - (\beta\mathcal{L}_M/2)^2} + \frac{1}{2} \sum_{j=1,2} I_j  \bigg\}
\end{align}
where
\begin{align}
    I_j &= \infint\D s \:\frac{e^{-s^2/4\sigma^2}e^{-i\Omega s}}{\sinh^2(\beta s/2-i\varepsilon) - (\beta \mathcal{L}_{L_j}/2)^2} \bigg\}
\end{align}
\end{widetext}
and we have defined the `mutual information distances' as 
\begin{align}
    \mathcal{L}_{L_1} &= 2R_D \sin \left( \frac{\theta_2' - \theta_1}{2} \right) \\
    \mathcal{L}_{L_2} &= 2R_D \sin \left( \frac{\theta_1' - \theta_2}{2} \right).
\end{align}
These quantities measure the distances between the `outer trajectories' (i.e.\ the first branch of detector $A$'s superposition, and the second branch of detector $B$'s superposition) and the `inner trajectories' (i.e.\ the second branch of detector $A$'s superposition, and the first branch of detector $B$'s superposition) of the detectors respectively. Note that for detectors on classical trajectories, the mutual information distance is exactly equal to the entangling distance; the only difference between the $\mathcal{L}$ and $\mathcal{M}$ terms in the bipartite detector density matrix is the time-ordering of the integrals in the latter. The Wightman functions on the other hand, are exactly the same in both terms.

When considering superpositions, the behaviour of the $\mathcal{L}$ term is altered significantly, since one has to account for the correlations between the field operators evaluated between four pairs of trajectories. In particular, the nonlocal correlation term does not straightforwardly decrease with the entangling distance, as the entangling term $\mathcal{M}$ does. As the entangling distance increases, the field correlations between certain pairs of trajectories may decrease; however for different pairs of trajectories, they may actually increase. To understand this, let us fix $\theta_1 = 0$, in which case $\mathcal{L}_{L_1}$ is strictly positive and increases monotonically with the entangling and superposition distances. This is not generally true for $\mathcal{L}_{L_2}$. For example, if one fixes $\theta_1'$ and we initially assume $\theta_1' > \theta_2$, then as the entangling distance increases, $\mathcal{L}_{L_2}$ changes from positive to negative values, and is zero when the second branch of detector $A$'s superposition overlaps with the first branch of detector $B$'s superposition. We emphasise that the correlation term $I_2$ is well-behaved at $\mathcal{L}_{L_2} = 0$ and its vicinity. That is, 
\begin{align}
    \lim_{\mathcal{L}_{L_2} \to 0} I_2 = \frac{\mathcal{P}_C}{4} &= \frac{\lambda^2\sqrt{\pi\sigma^2}}{4} \infint\D s \:\frac{e^{-s^2/4\sigma^2} e^{-i\Omega s}}{\sinh^2(\beta s/2-i\varepsilon) }   
\end{align}
is mathematically well-defined, where $\mathcal{P}_C$ is the transition probability of a single classical detector.

\begin{figure}[h]
    \centering
    \includegraphics[width=\linewidth]{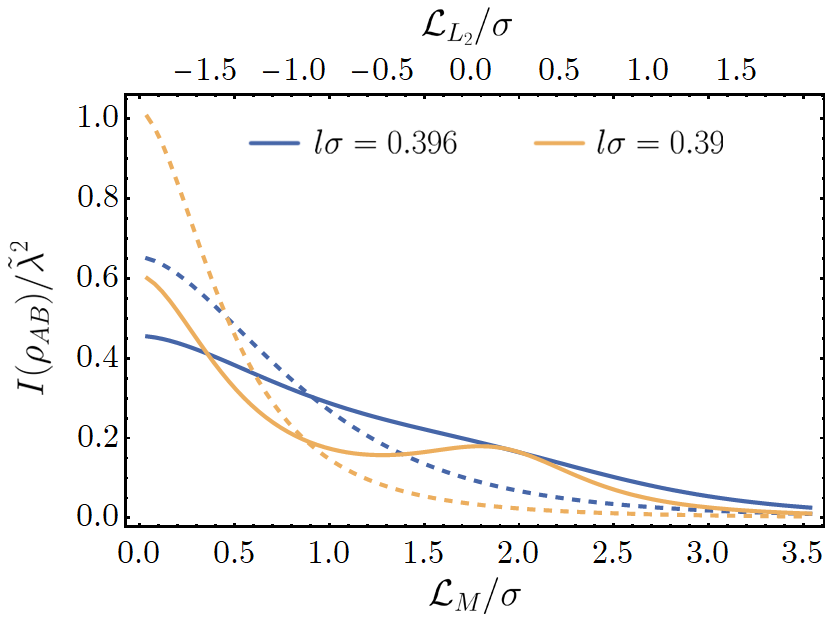}
    \caption{Mutual information, $I(\rho_{AB})/\tilde{\lambda}^2$, harvested by quantum-controlled detectors (solid) as a function of the entangling distance (bottom axis) $\mathcal{L}_M/\sigma$ or equivalently, the mutual information distance $\mathcal{L}_{L_2}/\sigma$, for (solid) detectors in superposition with $\mathcal{L}_S/\sigma = 1.91$ and (dashed) detectors on classical paths, where $\Omega \sigma = 0.02$ and $R_D/\sigma = 2.5$.}
    \label{fig:mutualdesitter2}
\end{figure}
\begin{figure}
    \centering
    \includegraphics[width=\linewidth]{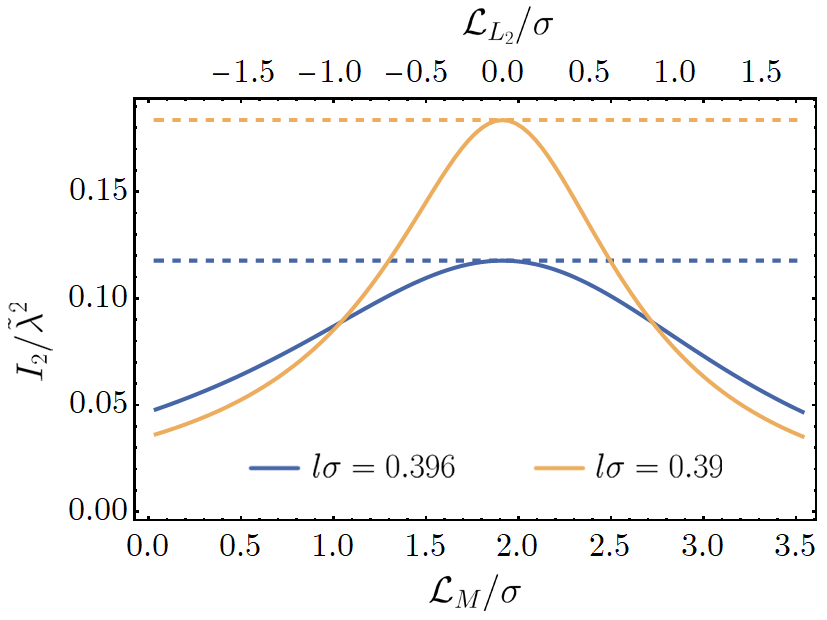}
    \caption{$I_2/\tilde{\lambda}^2$ as a function of the entangling distance, $\mathcal{L}_M/\sigma $ for the values corresponding to Fig.\ \ref{fig:mutualdesitter2}. The dashed lines correspond to $\mathcal{P}_C/4$, the transition probability of a single classical detector at the same $r$, $\phi$.}
    \label{fig:MIvsW}
\end{figure}
In Fig.\ \ref{fig:mutualdesitter2}, the mutual information between the detectors is plotted as a function of the entangling distance, for fixed $\theta_1$, $\theta_1'$. For detectors on classical trajectories, the mutual information monotonically decreases with increasing distance. This is also true for detectors in superposition, when the de Sitter length (curvature) is sufficiently small. For a sufficiently curved spacetime -- for detectors at fixed $r$, this means that the distance to the cosmological horizon decreases -- a turning point in the mutual information emerges, so that it begins to increase with the entangling distance, peaking at $\mathcal{L}_{L_2} = 0$, before decaying again as $\mathcal{L}_{L_2}$ becomes increasingly negative. This behaviour, as alluded to, can be traced to the $I_2$ term in the nonlocal detector correlations. For the initial setup of Fig.\ \ref{fig:mutualdesitter2}, the mutual information distance $\mathcal{L}_{L_2}$ initially decreases so that the field correlations between these two paths in the superposition grow, and are maximised when the trajectories overlap exactly. The reason this feature only appears for more highly curved spacetimes can be traced to the increased temperature of the field. Detectors at a fixed $r$ will perceive the field state to be populated with increasingly many thermal particles as the curvature $l$ of the spacetime increases, leading to a higher transition probability. When the superposed paths overlap exactly, $I_2$ is proportional to this transition probability, which maximises the total correlation between any two spacetime points (see Fig.\ \ref{fig:MIvsW}). In this case, the correlations are evaluated with respect to the proper times $\tau$, $\tau'$ along the single worldline. 

The enhancement of the mutual information in certain regimes highlights the interesting physical effects induced by quantum indeterminacy. In the classical trajectory case, the distance between spacetime points is a general indicator of the strength of the correlations, quantum or classical, between those points. By introducing a superposition of trajectory states, the distance between the two detectors is no longer well-defined. Hence, the correlations between them can seemingly increase as the detectors get further apart (that is if one naively uses the `classical' entangling distance, to quantify this). 
\begin{figure}[h]
    \centering
    \includegraphics[width=\linewidth]{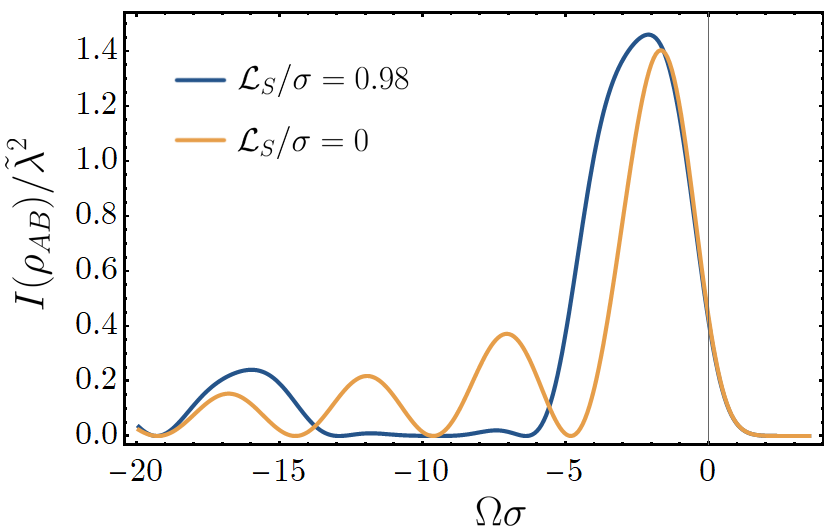}
    \caption{Mutual information, $I(\rho_{AB})/\tilde{\lambda}^2$ harvested by the detectors as a function of the energy gap, for $\mathcal{L}_S /\sigma = 0$ (orange) and $\mathcal{L}_S /\sigma = 0.98$ (blue). The other parameters we have used are $l\sigma = 0.1$, $R_D/\sigma = 2.5$. }
    \label{fig:MIvsWdesitter}
\end{figure}
\begin{figure}[h]
    \centering
    \includegraphics[width=\linewidth]{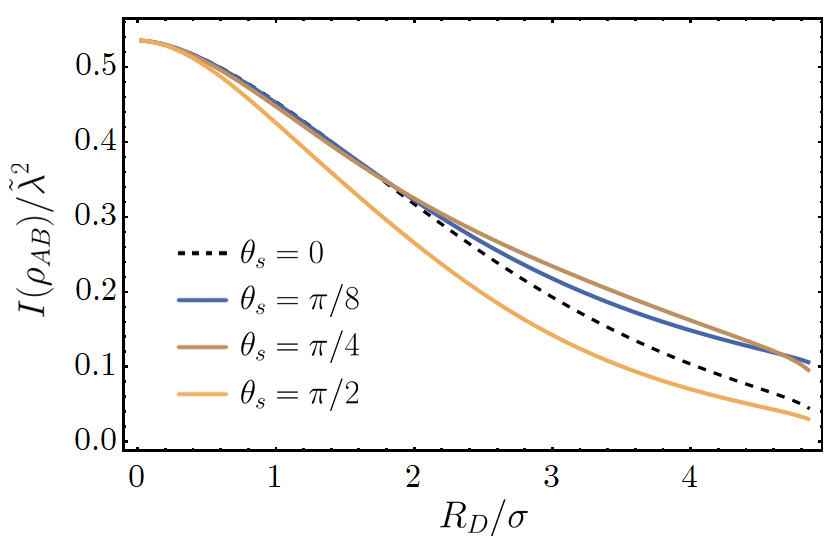}
    \caption{Mutual information harvested by the detectors on classical trajectories and in quantum-controlled superpositions, as a function of the radial coordinate. We have used the parameters $l \sigma = 0.2$, $\theta_M = \pi/6$, $\Omega\sigma = 0.02$.}
    \label{fig:mutualdesitter1}
\end{figure}

In Fig.\ \ref{fig:MIvsWdesitter}, the mutual information is plotted as a function of the energy gap, $\Omega\sigma$, showing its exponential suppression for positive energy gaps, and oscillatory behaviour for large negative energy gaps. Clearly, the amount of mutual information that can be harvested from the field (and whether this is more or less than detectors on classical trajectories) is highly sensitive to the value of $\Omega$. Finally in Fig.\ \ref{fig:mutualdesitter1}, we have plotted the mutual information harvested by the detectors as a function of the radial coordinate, $R_D/\sigma$. For the parameters shown, the mutual information decreases as the detectors approach the horizon. The rate of this decrease depends on the strength of the nonlocal correlations between the outer and inner trajectories of the quantum-controlled detectors.


\subsection{Temporal superpositions}
In contrast to entanglement harvesting, the mutual-correlation terms, $\mathcal{L}$, are identical for both the past-future and cause-effect superpositions. They are given by
\begin{align}
    \mathcal{L} &= \frac{\lambda^2\sqrt{\pi\sigma^2}}{2} \Bigg\{ \infint\D s \: \frac{e^{-s^2/4\sigma^2}e^{-i\Omega s}}{\sinh^2(\beta s/2-i\varepsilon) - (\beta\mathcal{L}_M /2)^2} \nonumber \\ 
    & + e^{-\tau_0^2/\sigma^2} \infint\D s \: \frac{e^{-s^2/4\sigma^2}e^{-i\Omega s}\cosh(s\tau_0/\sigma^2)}{\sinh^2(\beta s/2-i\varepsilon) - (\beta \mathcal{L}_L/2)^2} \Bigg\} 
\end{align}
where the local transition probabilities are identical to the expressions previously stated. 
\begin{figure}[h]
    \centering
    \includegraphics[width=\linewidth]{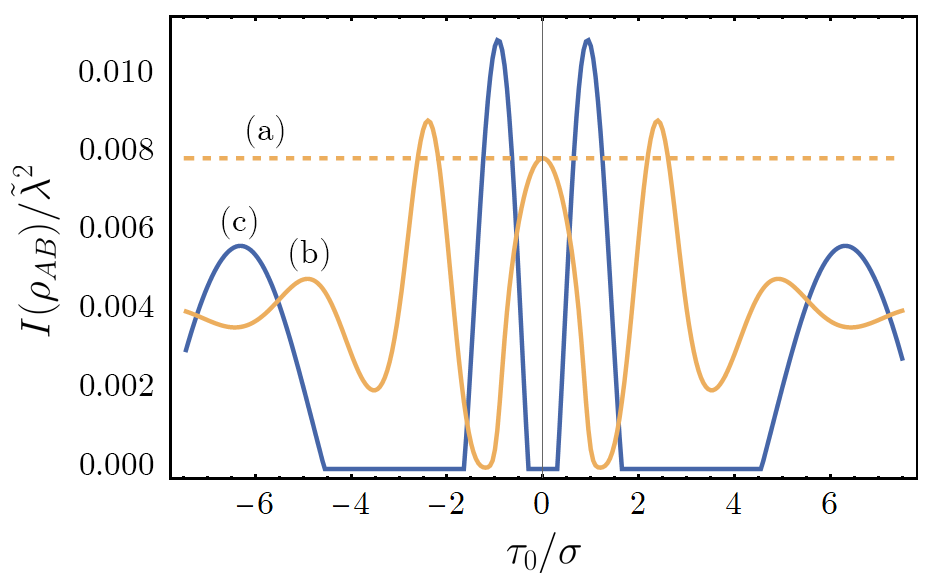}
    \caption{Plot of $I(\rho_{AB})/\tilde{\lambda}^2$ (b) and $\mathcal{C}_{AB}/4\tilde{\lambda}^2$ (c) harvested by detectors in a past-future superposition of temporal orders, as a function of $\tau_0/\sigma$. The dashed line (a) represents the mutual information harvested by detectors on classical trajectories (the \textit{entanglement} harvested by classical detectors is zero). We have used the parameters $l\sigma = 0.2$, $R_D/\sigma = 2.5$, $\Omega\sigma = 1$. }
    \label{fig:MItemporaldesitter}
\end{figure}
In Fig.\ \ref{fig:MItemporaldesitter}, we have plotted the mutual information (orange) as a function of the superposition time-delay, $\tau_0/\sigma$. The concurrence obtained for the past-future superposition (scaled by 1/4), is displayed in blue.
\begin{figure}[h]
    \centering
    \includegraphics[width=0.75\linewidth]{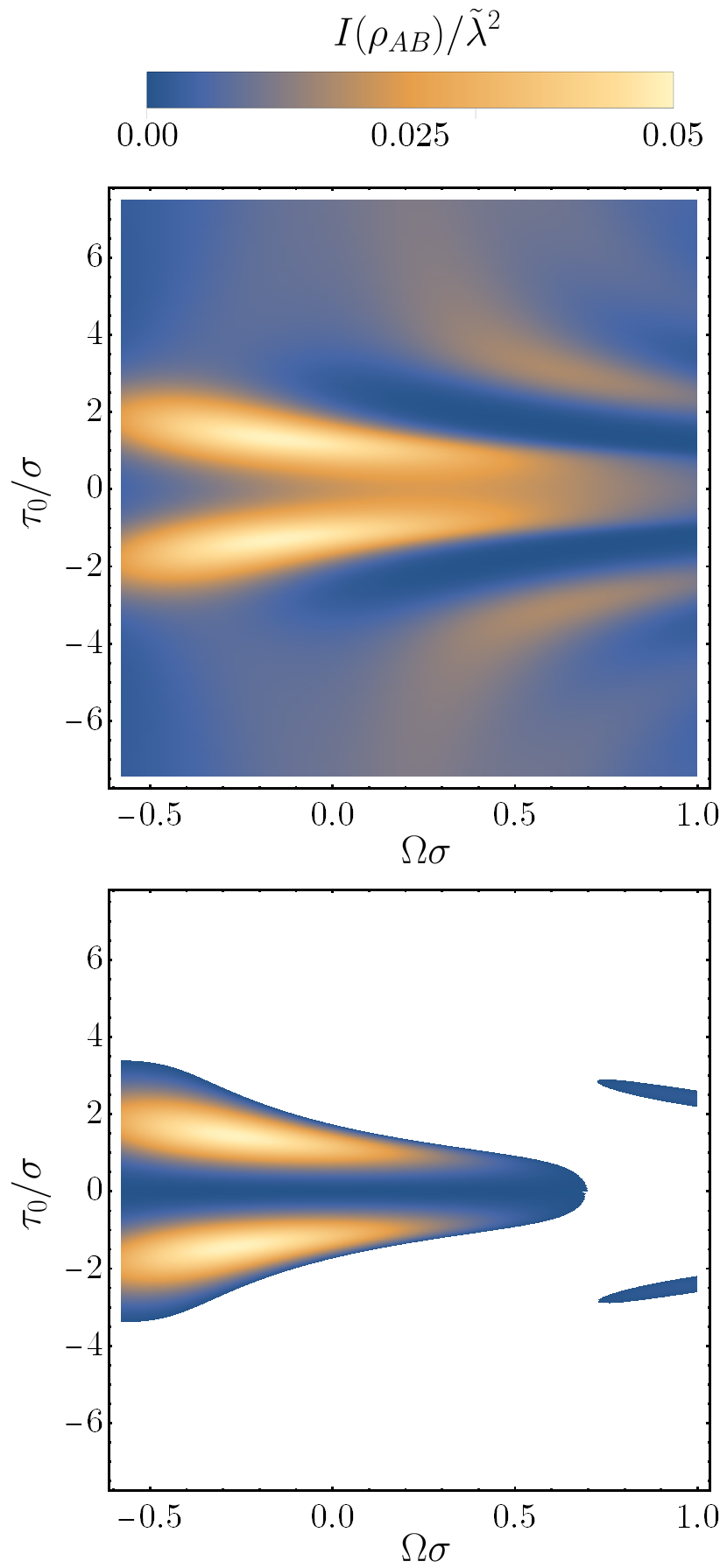}
    \caption{(top) Mutual information, $I(\rho_{AB})/\tilde{\lambda}^2$, harvested by detectors interacting with the field in a quantum-controlled temporal superposition, as a function of $\tau_0/\sigma$ and $\Omega\sigma$. We have used the parameters $l\sigma = 0.2$, $R_D/\sigma = 2.5$ and $\mathcal{L}_M/\sigma  = 3.53$. (bottom) The difference in the mutual information harvested by detectors in temporal superposition, compared with the analogous setup without the superposition. The coloured regions are those in which the superposed detectors harvest more mutual information than classical detectors, and vice versa for the white regions.}
    \label{fig:densityMIdesittertemporal1}
\end{figure}
We find that the mutual information oscillates with $\tau_0/\sigma$, eventually decaying to an equilibrium value for $|\tau_0/\sigma|\gg 1$. Interestingly, the mutual information peaks in regions where the entanglement vanishes, for example at $\tau_0/\sigma = 0$ and $|\tau_0/\sigma| \simeq 2.5$. Contrarily, we also discover that the troughs in the mutual information occur at the peaks of the concurrence, for example at $|\tau_0 /\sigma | \simeq 1.5$. This oscillatory behaviour indicates that the interference effects experienced by the detectors are highly sensitive to the precise field regions probed by the interaction. In particular in the regime $|\tau_0/\sigma | \lesssim 5$, the correlations between the detector approximately swap between mutual information and entanglement.

\begin{figure}[h]
    \centering
    \includegraphics[width=0.75\linewidth]{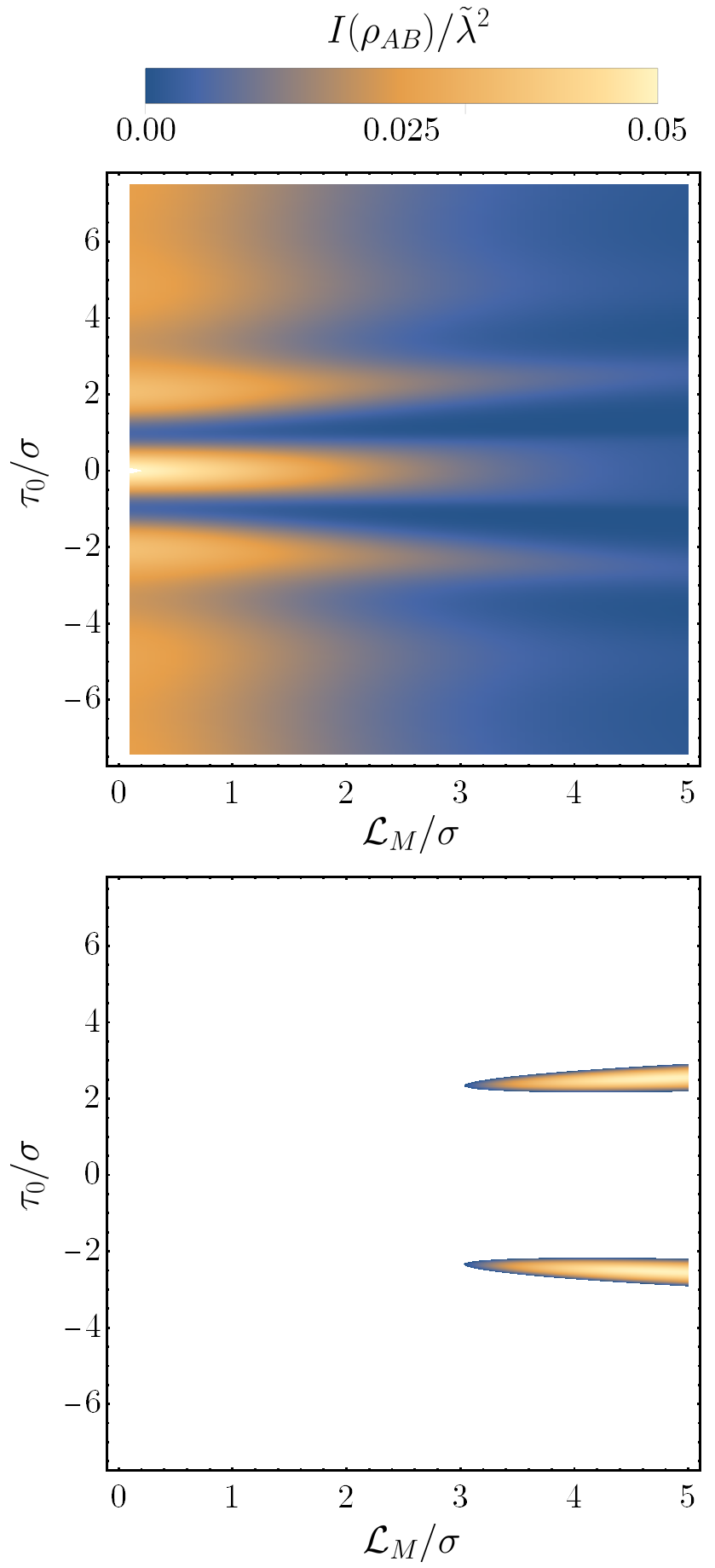}
    \caption{(top) Mutual information, $I(\rho_{AB})/\tilde{\lambda}^2$, harvested by detectors in a quantum-controlled temporal superposition, as a function of $\mathcal{L}_M/\sigma$ and $\tau_0/\sigma$. We have used the parameters $l\sigma = 0.2$, $R_D/\sigma = 2.5$ and $\mathcal{L}_M/\sigma  = 3.53$. (bottom) The difference in the mutual information harvested by detectors in temporal superposition, compared with the analogous setup without the superposition. The coloured regions show those in which superposed detectors harvest more mutual information than their classical counterparts. }
    \label{fig:densityMIdesittertemporal2}
\end{figure}

In Fig.\ \ref{fig:densityMIdesittertemporal1}, we have plotted the mutual information as a function of the energy gap, $\Omega\sigma$, and the time-delay, $\tau_0/\sigma$. The presence of the temporal superposition enhances mutual information harvesting in the $|\tau_0/\sigma| \ll 1$ and negative energy gap regions, compared to detectors interacting with the field in classical spacetime regions. Similarly in Fig.\ \ref{fig:densityMIdesittertemporal2}, we have plotted the mutual information as a function of the entangling distance, $\mathcal{L}_M/\sigma$, and the time-delay. Only at large separations does the presence of the superposition enhance mutual information harvesting, over and against detectors on classical spacetime trajectories.  

\subsection{Measuring the control in an arbitrary superposition state}
We have thus far considered scenarios where the control system is measured in the same state as it was initially prepared in. More generally, we perform the measurement in some arbitrary superposition basis, given by 
\begin{align}
    |\chi\rangle &= \frac{1}{\sqrt{N}} \sum_{i=1}^N e^{-i\varphi_i }|i\rangle_C  
\end{align}
where $\varphi_i \in \mathbb{R}$, so that the final detector-field state is given by 
\begin{align}
    |\Psi\rangle_{FD} &= \frac{1}{N} \sum_{i=1}^N  e^{i\varphi_i} \hat{U}_i |0\rangle  |g\rangle_A |g\rangle_B 
\end{align}
and $\hat{U}_i$ is defined in Eq.\ (\ref{trajectoryunitary}). 


For detectors in a superposition of two trajectories, the local transition probability and entangling term are given respectively by 
\begin{align}
    \mathcal{P}_D &= \frac{\lambda^2\sqrt{\pi\sigma^2}}{2} \Bigg\{ \int\D s \: \frac{e^{-s^2/4\sigma^2}e^{-i\Omega s}}{\sinh^2(\beta s/2-i\varepsilon)} \nonumber \\
    & + \cos(\varphi_1 - \varphi_2 ) \int\D s \: \frac{e^{-s^2/4\sigma^2}e^{-i\Omega s}}{\sinh^2(\beta s/2-i\varepsilon - (\beta \mathcal{L}_S /2)^2} \Bigg\} 
    \intertext{and}
    \mathcal{M} &= \lambda^2\sqrt{\pi\sigma^2} \int_0^\infty \D s \:\frac{ e^{-s^2/4\sigma^2}\big( 1 + \cos ( \varphi_1 - \varphi_2 ) \big) }{\sinh^2(\beta s/2-i\varepsilon) - (\beta \mathcal{L}_M/2)^2} .
\end{align}
For brevity, we consider only the entanglement harvested by the detectors in this scenario. We notice immediately that if the final state of the control is orthogonal to the initial state, namely
\begin{align}
    |\chi_-\rangle &= \frac{1}{\sqrt{2}} \big( |1 \rangle_C - |2\rangle_C \big) 
\end{align}
then $\mathcal{M} = 0$ (i.e\ $\varphi_1 - \varphi_2 = \pi$), implying that it is impossible for the detectors to harvest entanglement. 


\begin{figure}[h]
    \centering
    \includegraphics[width=0.75\linewidth]{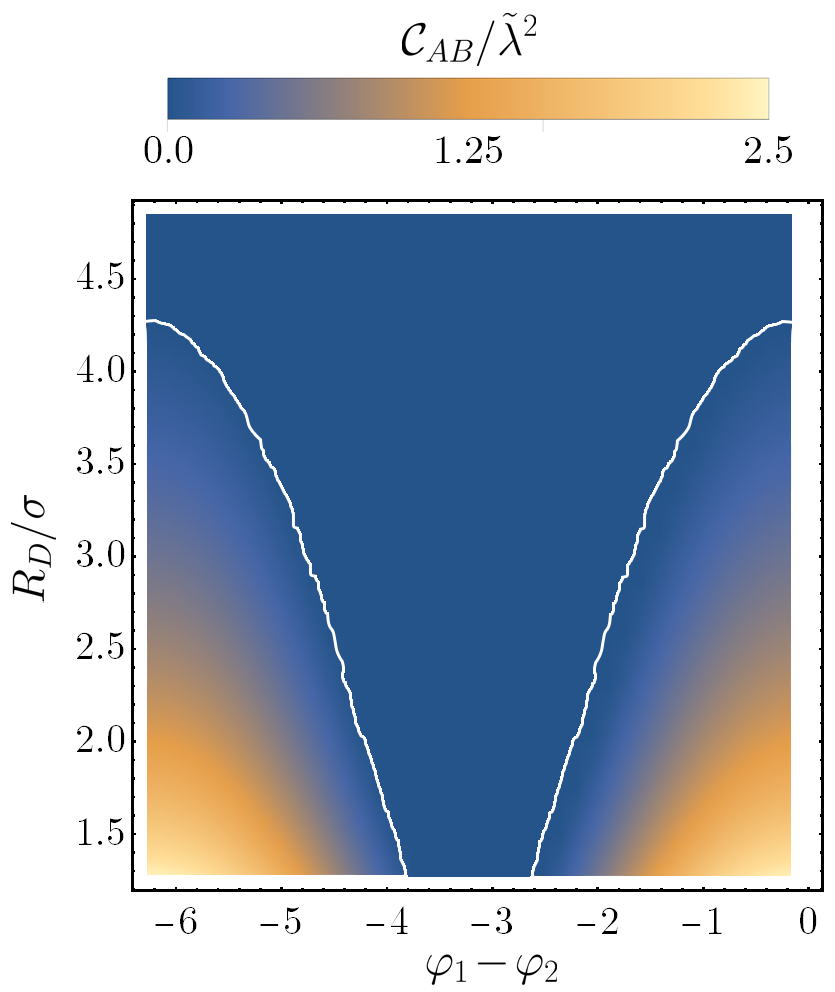}
    \caption{Concurrence, $\mathcal{C}_{AB}/\tilde{\lambda}^2$, between the detectors, as a function of the relative phase $\varphi_1 - \varphi_2$ and the radial coordinate, $R_D/\sigma$. We have used $l\sigma = 0.2$, $\Omega\sigma = 0.02$, $\theta_M = \pi/6$ and $\theta_S = \pi/2$. The white line shows the contour of zero concurrence. }
    \label{fig:relativephasedesitter}
\end{figure}
In Fig.\ \ref{fig:relativephasedesitter}, we have plotted the concurrence between the two detectors as a function of the relative phase $\varphi_1 - \varphi_2$, and the radial coordinate, $R_D/\sigma$. We find that the concurrence is maximised when the control is measured in its initial state; that is, $\varphi_1 - \varphi_2 = 2\pi n$ where $n \in\mathbb{Z}$. As noted above, for a relative phase 
of $\pi$ (and values near this) between the control states, the concurrence vanishes completely.

\section{Finite-temperature entanglement harvesting in Minkowski Spacetime}\label{thermalMinkowski}
We now consider entanglement harvesting with finite-temperature fields in flat Minkowski spacetime. It is well-known that the single detectors on classical trajectories respond identically to Gibbons-Hawking radiation in the de Sitter conformal vacuum and thermal radiation in flat Minkowski spacetime. It was shown by Ver Steeg and Menicucci \cite{ver2009entangling} that by introducing a second detector, these spacetimes can in principle be distinguished through the amount of harvested entanglement, when their separation is larger than the characteristic length scale of the de Sitter expansion. More recently, it was shown that a single quantum-controlled detector can discern the global properties of these spacetimes \cite{foo2020thermality}, allowing it to distinguish between de Sitter expansion and thermal radiation in Minkowski spacetime, through a measurement of its transition probability. Other studies of entanglement dynamics and degradation in these spacetimes \cite{tian2013geometric,tian2014dynamics,tian2016detecting,akhtar2019open} employed the resonance Casimir-Polder interaction and an open quantum system formalism. 

In the following, we consider two UdW detectors, each prepared in a superposition of static trajectories separated by the superposition distance $\mathcal{L}_S$, with the $i$th trajectories each separated by the entangling distance, $\mathcal{L}_M$. The Wightman functions evaluated along the individual trajectories are identical to the de Sitter case, which means that single detectors on classical trajectories respond identically to the field in both these spacetimes (with the identification $\beta \Leftrightarrow \kappa$). However the non-local Wightman functions take a slightly different form, and are given by \cite{weldon2000thermal}
\begin{align}\label{thermal1}
    \mathcal{W}_\text{th-$i$}(s) &= \frac{\kappa \big( \coth  \frac{\kappa }{2}(\mathcal{L}_S-s')  + \coth \frac{\kappa }{2} (\mathcal{L}_S+s' )  \big)}{16\pi^2\mathcal{L}_S}  \\ \label{thermal3}
    \mathcal{W}_\text{th-$m$}(s) &= \frac{\kappa \big( \coth  \frac{\kappa }{2}(\mathcal{L}_M-s')  + \coth \frac{\kappa }{2} (\mathcal{L}_M+s' )  \big)}{16\pi^2\mathcal{L}_M}  
\end{align}
where we have defined $s' = s - i\varepsilon$. 

\subsection{Spatial superpositions}
\begin{figure}[h]
    \centering
    \includegraphics[width=\linewidth]{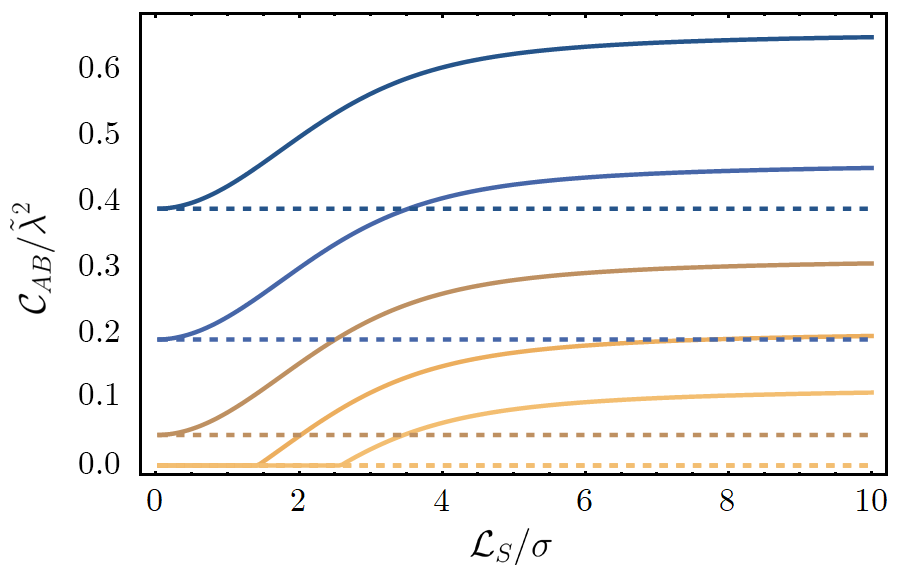}
    \caption{Concurrence, $\mathcal{C}_{AB}/\tilde{\lambda}^2$, harvested by the detectors, as a function of $\mathcal{L}_S/\sigma$ for $\mathcal{L}_M/\sigma = 1.25$, $1.5$, $1.75$, $2$, $2.25$ (from top to bottom). The dashed lines represent the concurrence between detectors on classical trajectories. We have used the parameters $l\sigma = 0.2$, $\Omega\sigma = 0.2$,  }
    \label{fig:thermalspatial2}
\end{figure}

The expressions for the local excitation probabilities, $\mathcal{P}_D$, and the entangling term, $\mathcal{M}$, take the same form as those used for the de Sitter case, apart from the relevant Wightman functions. In Fig.\ \ref{fig:thermalspatial2}, we have plotted the concurrence between the detectors as a function of the superposition distance, $\mathcal{L}_S$, for different values of $\kappa\sigma$. 
One immediately notices the similarity between Fig.\ \ref{fig:thermalspatial2} and Fig.\ \ref{fig:deistterspatialLS} whereby the concurrence increases in an asymptotic fashion as the superposition distance increases.  Again, we obtain this result because at large superposition distances, the interference terms in the detector transition probabilities, $\mathcal{P}_{ij,D}$, vanish (i.e.\ the correlations between the superposed interaction regions of each detector decay with the superposition distance). Note especially that the contributions to $\mathcal{P}_D$ from the individual trajectories of the superposition are independent of $\mathcal{L}_S$, unlike the interference terms. This further emphasises the phenomenon of enhanced entanglement harvesting in the presence of superposition, and we conjecture that it is also true for more generic scenarios. 

\begin{figure}[h]
    \centering
    \includegraphics[width=\linewidth]{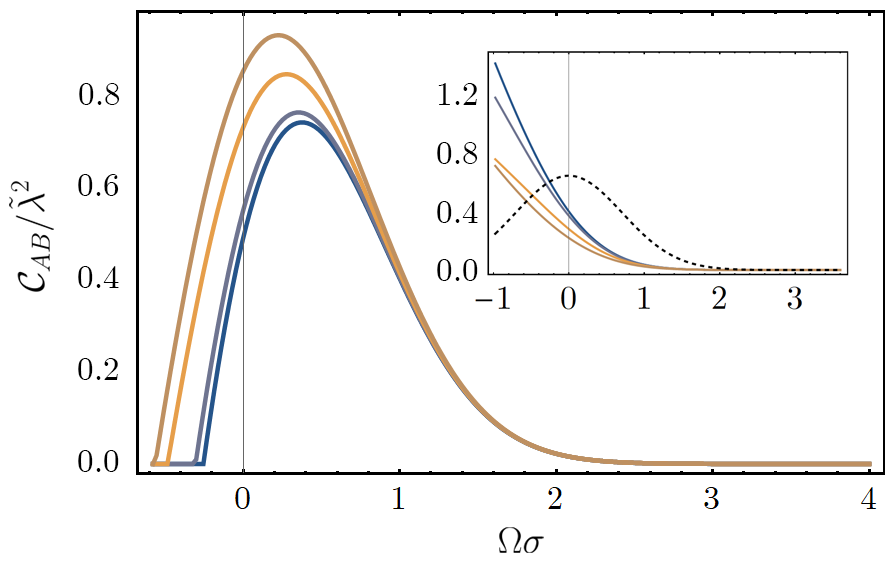}
    \caption{Concurrence, $\mathcal{C}_{AB}$, as a function of the energy gap, $\Omega\sigma$, for $\mathcal{L}_S/\sigma = 0$, $1$, $2.5$, $5$ (top to bottom). The inset shows the local noise terms against the entangling term, $\mathcal{M}$ (dashed line). The other parameters we have used are $\mathcal{L}_M /\sigma = 1$ and $\kappa\sigma = 0.2$. }
    \label{fig:thermalspatial1}
\end{figure}

In Fig.\ \ref{fig:thermalspatial1}, we have plotted the concurrence harvested by the detectors as a function of the energy gap, $\Omega\sigma$. We find that entanglement harvesting is generally inhibited for negative energy gaps, corresponding to detectors initialised in their excited state. This closely resembles the behaviour of the concurrence for detectors in de Sitter spacetime. Although the entangling term is symmetric with respect to $\Omega$, the transition probability grows unbounded as $\Omega\sigma$ becomes more negative (see the inset of Fig.\ \ref{fig:thermalspatial1}). Again, we notice that the concurrence between the detectors is amplified for detectors in spatial superpositions, and most notably so for small energy gaps, $|\Omega\sigma| \ll 1$. 
\begin{figure}[h]
    \centering
    \includegraphics[width=\linewidth]{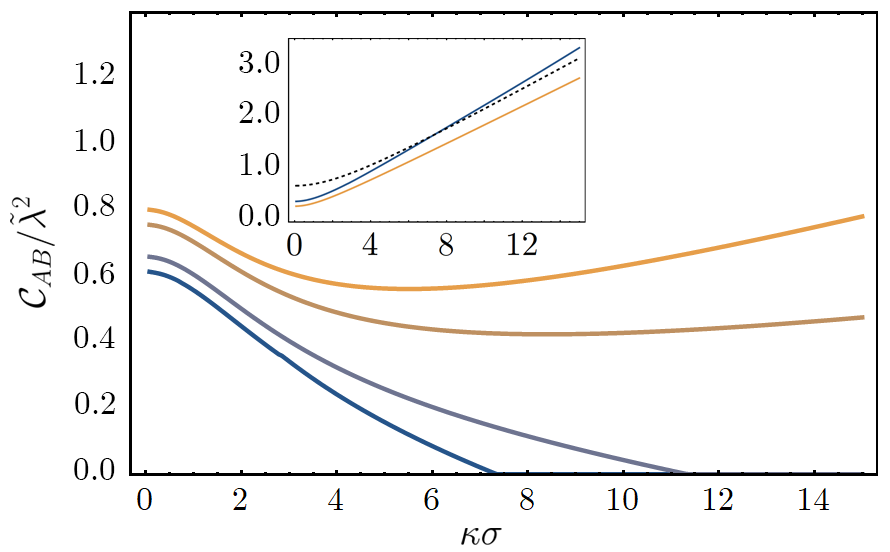}
    \caption{Concurrence, $\mathcal{C}_{AB}$ as a function of $\kappa\sigma$ where we have plotted $\mathcal{L}_S/\sigma = 0, 1,2,2.5$ top to bottom (orange to blue lines). The inset compares the local noise term against the entangling term (dashed line) for $\mathcal{L}_S/\sigma = 0, 2.5$ top to bottom (blue and orange plots, respectively). The other parameters we have used are $\mathcal{L}_M/\sigma = 1$, $\Omega\sigma = 0.1$. }
    \label{fig:thermalspatial3}
\end{figure}

In Fig.\ \ref{fig:thermalspatial3}, we have plotted the concurrence as a function of $\kappa\sigma$, comparing detectors travelling on classical trajectories with those in quantum-controlled spatial superpositions. For sufficiently small superposition distances, the concurrence decays with increasing temperature, as the noise induced by the presence of thermal particles dominates over the correlations between the two detectors. Nevertheless, there is still an advantage of using detectors in superposition, due to the relative suppression of the local noise terms. Quite remarkably, we observe that for sufficiently large superposition distances $(\mathcal{L}_S/\sigma\gg 1)$, and temperatures, $(\kappa\sigma\gg 1)$,
the concurrence reaches a global minima and grows with increasing temperature.

To understand this, note that although $\mathcal{M}$ grows with increasing temperature, and for detectors on classical trajectories, it is eventually overtaken by the local noise term, $\mathcal{P}_D$ for small  superposition distances. 
However for sufficiently large superposition distances, the
interference terms die out,
and the  transition probability
approaches half its value attained for a classical trajectory.
This has the effect of suppressing the noise terms below the entangling term, even for large values of $\kappa \sigma$.

These terms grow linearly for large $\kappa \sigma$, which leads us to the counter-intuitive conclusion, that  (to second-order perturbation theory) for detectors in sufficiently distant spatial superpositions, the amount of entanglement harvested from the field grows with increasing temperature. 
\begin{figure}
    \centering
    \includegraphics[width=0.8\linewidth]{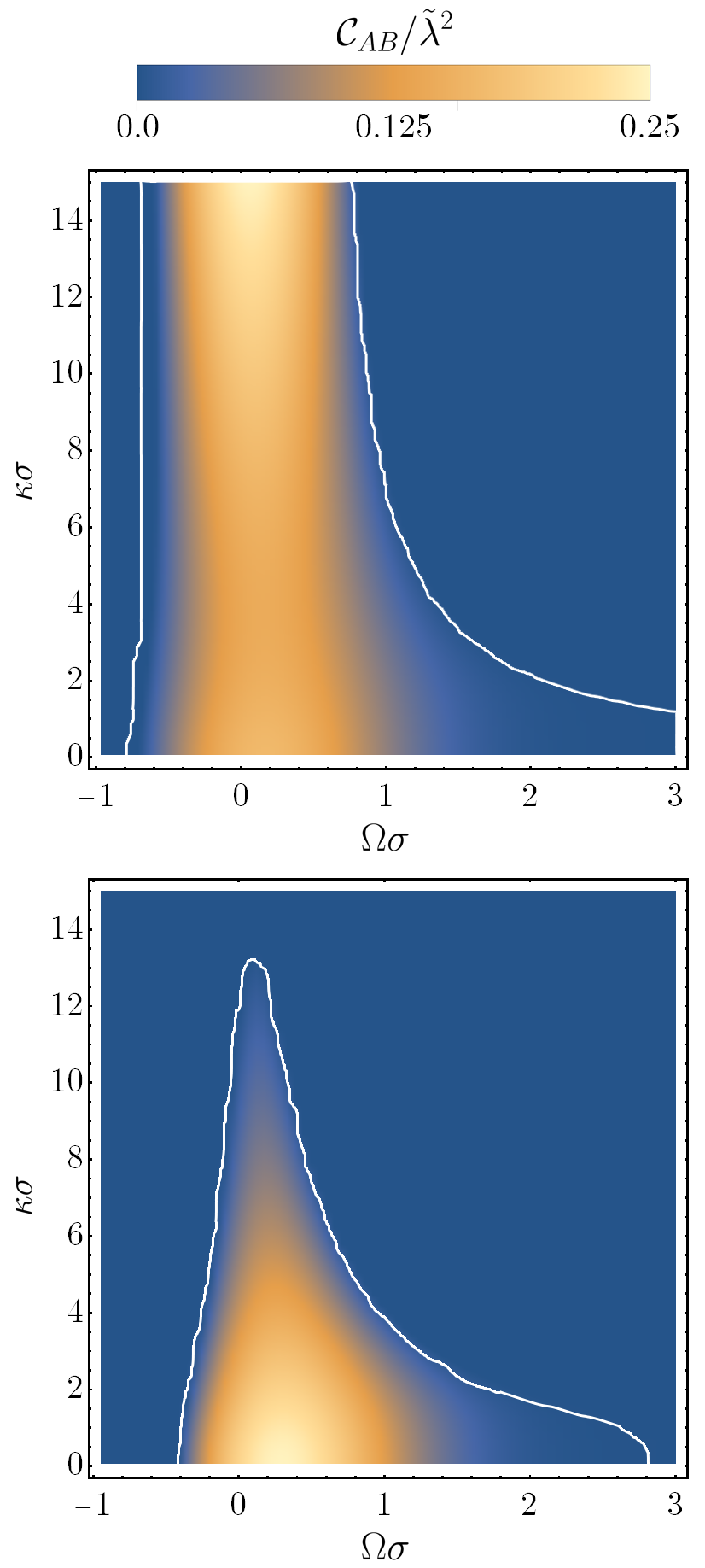}
    \caption{(top) Concurrence, $\mathcal{C}_{AB}/\tilde{\lambda}^2$, between the detectors in a spatial superposition compared with (bottom) detectors on classical trajectories. We have used the $\Omega\sigma = 0.1$, $\mathcal{L}_M/\sigma = 1.5$ and in the top figure, a superposition distance of $\mathcal{L}_S /\sigma = 5$. The white lines represent the contour of zero entanglement. }
    \label{fig:thermaldensity1}
\end{figure}
This result demonstrates (to our knowledge) the first violation of the no-go theorem derived by Simidzija in \cite{simidzijathermalstates}, where it was shown that for a thermal field state, (a) the amount of entanglement harvested by identical UdW detectors decreases with the temperature, and (b) that there always exists a threshold temperature above which identical UdW detectors cannot harvest any entanglement. This theorem was derived in the regime of perturbation theory for detectors with arbitrary switching functions and spatial profiles, and applies regardless of the dimension of the spacetime or the mass of the field \cite{simidzijathermalstates}. Our results reveal that the theorem relies on an additional tacit assumption that trajectories of the detectors are classical.  

The density plots of Fig.\ \ref{fig:thermaldensity1}  further illustrate the difference between detectors on superposed trajectories and those on classical paths. Clearly, the detectors in superposition can become entangled in a significantly broader region of the parameter space, and for sufficiently small energy gaps, $|\Omega\sigma|\ll 1$, the concurrence grows with the temperature of the field.  Note especially that the detector transition probability $\mathcal{P}_{D}$ (including the constituent `local' and `interference' contributions) behaves as we should expect with increasing temperature; that is, it increases monotonically with $\kappa \sigma$. In other words  higher temperatures amplify excitations in the superposed detector, which would typically inhibit entanglement harvesting in a classical detector. However alongside the amplification of excitations with increasing temperature is the \textit{suppression} of excitations with increasing superposition distance, $\mathcal{L}_S$. For sufficiently large $\mathcal{L}_S$, this suppression is such that $\mathcal{P}_D$ is always smaller than $\mathcal{M}$, term even for higher and increasing temperatures. 

Due to the unbounded growth in these quantities,  perturbation theory will eventually break down.  It would be interesting to see how higher-order terms in $\lambda$ modify the above results. 

\section{Mutual Information Harvesting in Thermal Minkowski Spacetime}
We finally study the mutual information harvested by detectors in thermal Minkowski spacetime. The expressions for the $\mathcal{L}$ term are identical to those derived for the de Sitter scenario, with a replacement of the relevant Wightman functions. In particular, we define the mutual information distances
\begin{align}
    \mathcal{L}_{L_1} &= \mathcal{L}_S + \mathcal{L}_M \\
    \mathcal{L}_{L_2} &= \mathcal{L}_S - \mathcal{L}_M 
\end{align}
so that the $\mathcal{L}$ term in the detector density matrix becomes
\begin{widetext}
\begin{align}\label{lterm78}
    \mathcal{L} &= \frac{\kappa\lambda^2 \sigma}{32\pi^{3/2}} \Bigg\{ \frac{1}{\mathcal{L}_S}\infint \D s \: e^{-s^2/4\sigma^2}e^{-i\Omega s} \left( \coth \left( \frac{\kappa}{2}( \mathcal{L}_S - s' ) \right) + \coth \left( \frac{\kappa}{2} ( \mathcal{L}_S + s ' \right) \right) + \frac{1}{2} \sum_{j=1,2} I_j \bigg\}
\end{align}
where
\begin{align}
    I_j &= \frac{1}{\mathcal{L}_{L_j}} \infint\D s \:e^{-s^2/4\sigma^2 } e^{-i\Omega s } \left( \coth \left( \frac{\kappa}{2} ( \mathcal{L}_{L_j} - s' ) \right) + \coth \left( \frac{\kappa}{2} (\mathcal{L}_{L_j} + s' ) \right) \right) 
\end{align}
\end{widetext}
As with the de Sitter case, the integral $I_2$ in Eq.\ (\ref{lterm78}) reduces to (a quarter of) the transition probability of a single detector on a classical worldline in the thermal bath, as $\mathcal{L}_{L_2} \to 0$. 

\begin{figure}[h]
    \centering
    \includegraphics[width=\linewidth]{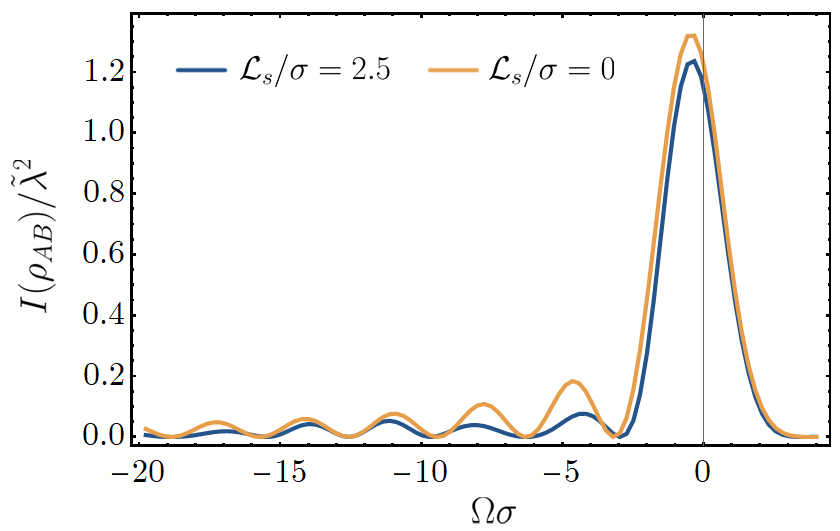}
    \caption{ Mutual information, $I(\rho_{AB})/\tilde{\lambda}^2$, harvested by the detectors as a function of the energy gap, $\Omega\sigma$. The other parameters used are $\mathcal{L}_M /\sigma =1$, $\kappa\sigma = 5$.} 
    \label{fig:MIthermalW}
\end{figure}

\begin{figure}[h]
    \centering
    \includegraphics[width=0.8\linewidth]{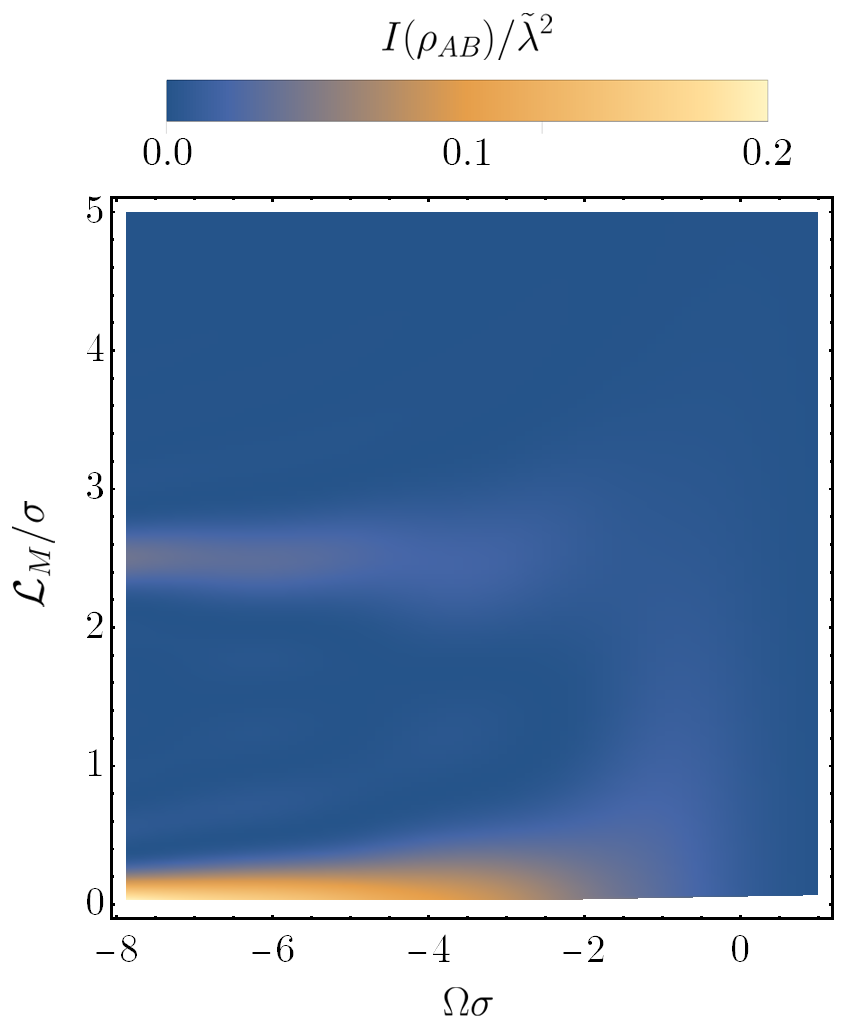}
    \caption{Mutual information, $I(\rho_{AB})/\tilde{\lambda}^2$, harvested by the detectors in a spatial superposition of trajectories as a function of $\Omega\sigma$ and $\mathcal{L}_M/\sigma$. In this plot, we have used the parameter settings $ l\sigma = 0.2$ and $\mathcal{L}_S/\sigma = 2.5$.}
    \label{fig:MIthermalLMW}
\end{figure}

Fig.\ \ref{fig:MIthermalW} displays the mutual information harvested by the detectors at fixed entangling distance, as a function of $\Omega\sigma$. The mutual information behaves in a similar manner to that already observed for the detectors in de Sitter spacetime, decaying in an oscillatory manner for large negative energy gaps. 
\begin{figure}[h]
    \centering
    \includegraphics[width=\linewidth]{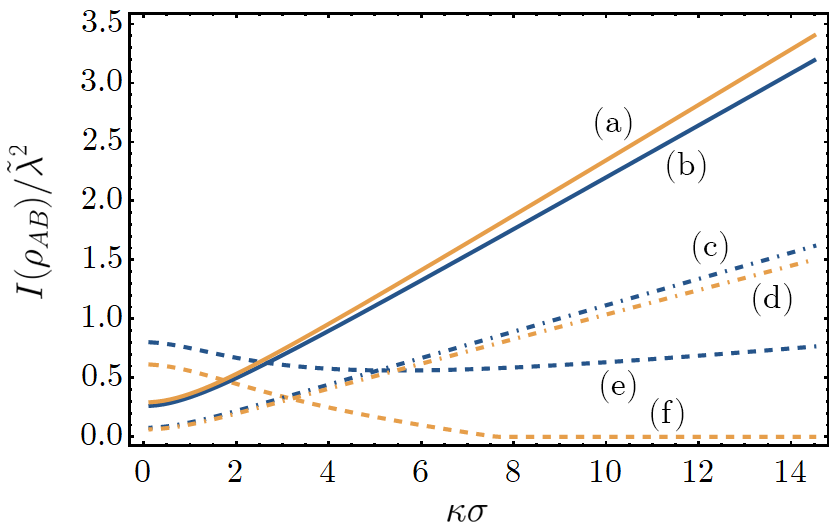}
    \caption{ Plot of the mutual information, $I(\rho_{AB})/\tilde{\lambda}^2$ as a function of $\kappa\sigma$, for (a) 
    $\mathcal{L}_S/\sigma = 2.5$, $\mathcal{L}_M/\sigma = 1.0$, (b)  $\mathcal{L}_S/\sigma = 0$, $\mathcal{L}_M/\sigma = 1.0$, (c) $\mathcal{L}_S/\sigma = 1.0$, $\mathcal{L}_M /\sigma =2.5$, and (d) $\mathcal{L}_S/\sigma = 0$, $\mathcal{L}_M/\sigma = 2.5$. In all plots, we have fixed $\Omega\sigma = 0.1$. Plots (e) and (f) show the concurrence, $\mathcal{C}_{AB}/\tilde{\lambda}^2$, corresponding to the setup of plots (a) and (b) respectively.}
    \label{fig:MIthermalA}
\end{figure}
In Fig.\ \ref{fig:MIthermalA}, we have plotted the mutual information as a function of $\kappa\sigma$. We find that the mutual information between the superposed detectors grows with the temperature of the field,   corroborating the result of \cite{simidzijathermalstates}. Note especially that for the detector setup plotted in Fig.\ \ref{fig:MIthermalA}(a) and (b), both the entanglement \textit{and} the mutual information grow with the temperature of the field at sufficiently high temperatures. Whether the superposition enhances or suppresses mutual information with respect to the classical-trajectory setup depends on the mutual information distances and the entangling distance (compared Fig.\ \ref{fig:MIthermalA}(a) and (b) with Fig.\ \ref{fig:MIthermalA}(c) and (d)).
\begin{figure}[h]
    \centering
    \includegraphics[width=\linewidth]{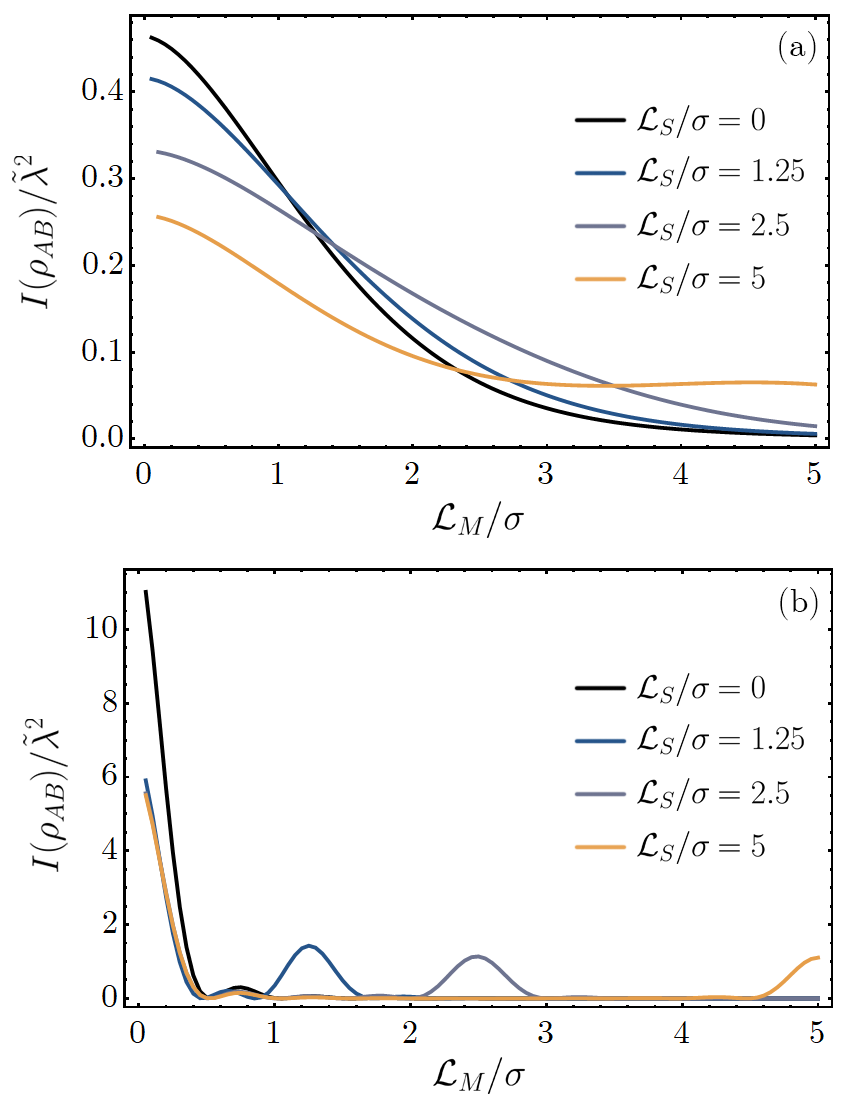}
    \caption{Mutual information, $I(\rho_{AB})/\tilde{\lambda}^2$, harvested by the detectors, (a) $\Omega \sigma = 0.1$ and (b) $\Omega \sigma = -6.0$. The other parameters we have used are $\kappa\sigma = 0.2$. The values of $\mathcal{L}_S/\sigma$ make sense of the peaks in the mutual information, namely at $\mathcal{L}_{L_2}/\sigma = 0$ for each detector setup.} 
    \label{fig:MIthermalLM}
\end{figure}
In Fig.\ \ref{fig:MIthermalLM}, the mutual information is plotted as a function of the entangling distance, $\mathcal{L}_M/\sigma$, for fixed values of the superposition distance, energy gap and temperature. The peak in the mutual information at the crossing of the superposed trajectories ($\mathcal{L}_{L_2} = 0$) is less noticeable for positive energy gaps than the de Sitter case, however still becomes manifest for negative energy gaps. This is also shown in Fig.\ \ref{fig:MIthermalLMW}, where the mutual information between the detectors is plotted as a function of the entangling distance and energy gap for a fixed superposition distance. Notably, for large negative energy gaps and small entangling distances, the mutual information is significantly enhanced. 

\section{Conclusion}\label{sec:conc}
It was already shown in \cite{foo2020unruhdewitt, Barbado:2020snx, foo2020thermality}, that the superposition of an UdW detector's path enables it to probe the global features of spacetime, including the background curvature and the causal relationship between the regions probed. In this paper, we have further demonstrated   that superposition enhances the ability of such detectors to extract entanglement from the quantum field. This is possible because the detectors interact with the field in a spatiotemporally delocalised fashion, and the resulting superposition of the interaction regions introduces interference effects which suppress the local noise perceived by each detector. Our results extend those presented in \cite{henderson2020quantum} which 
studied entanglement harvesting with indefinite causal order (i.e.\ superpositions of interaction times with the field); here we have studied the effects of both temporal and spatial superpositions, the mutual information between detectors in addition to their entanglement, and investigated curved spacetime and finite-temperature scenarios.

Although entanglement harvesting has been studied extensively in the literature, the quantum-controlled detector model opens the way for straightforward extensions to nearly all conceivable settings already considered. Already the present results strongly suggest that the effects of delocalised detector trajectories may elicit further insight into foundational questions about the entanglement structure of quantum fields in relativistic settings and curved spacetime. Furthermore, our work has direct relevance to the burgeoning field of quantum information and communication and quantum thermodynamics in the presence of quantum control of different channels; the quantum-controlled UdW detector model provides a field-theoretic realisation of such scenarios. 
\\\\
\acknowledgments 

The authors thank an anonymous referee who pointed out an error in the original manuscript. J.F.\ acknowledges support from the Australian Research Council Centre of Excellence for Quantum Computation and Communication Technology (Project
No.\ CE170100012). M.Z. acknowledges support from DECRA Grant DE180101443. R.B.M acknowledges support from the Natural Sciences and Engineering Research Council of Canada and from AOARD Grant FA2386-19-1-4077. 

\bibliography{main}

\end{document}